%% file: DSPACE_version.tex
\documentclass[sigconf,nonacm]{acmart}

\usepackage{graphicx}
\usepackage{tikz}
\usepackage{pgfplots}
\usepackage{pgfplotstable}
\usepgfplotslibrary{dateplot}
\pgfplotsset{compat=newest}
\usepgfplotslibrary{fillbetween}
\usetikzlibrary{shapes.geometric}
\pgfplotsset{colormap/hot}
\usepackage{multirow}
\usepackage{float}

\usepackage{siunitx}

\usepackage{placeins}

\usepackage[inline]{enumitem}

\usepackage{xspace}

\usepackage{csquotes}
\usepackage{framed}

\usepackage{comment}

\usepackage[nolist]{acronym}

\usepackage{soul}

\newboolean{showcomments}
\setboolean{showcomments}{false}
\usepackage{xurl}

\usepackage{cleveref}

\usepackage{nth}

\ifthenelse{\boolean{showcomments}}
{ \newcommand{\mynote}[3]{
		\protect\fbox{\bfseries\sffamily\scriptsize#1}~%
		{\small\textsf{\emph{\color{#3}{#2}}}}}}
{ \newcommand{\mynote}[3]{}\ignorespaces}

\newcommand{\lb}[1]{\mynote{Leo}{#1}{blue}}

\newcommand{\sd}[1]{\mynote{Sergio}{#1}{orange}}

\newcommand{\cf}{cf.\@\xspace}

\newcommand{\etal}{\textit{et al.}\@\xspace}
\newcommand{\eg}{\textit{e.g.}\@\xspace}

\newcommand{\ie}{\textit{i.e.}\@\xspace}

\newcommand{\lptp}{LibP2P\xspace}
\newcommand\para[1]{\vspace{0.05in} \noindent \textbf{#1.}}
\begin{document}
\input{acronyms}

% TODO authors
% \begin{comment}
% %%
% %% The "author" command and its associated commands are used to define
% %% the authors and their affiliations.
% %% Of note is the shared affiliation of the first two authors, and the
% %% "authornote" and "authornotemark" commands
% %% used to denote shared contribution to the research.
% \affiliation{%

% \affiliation{%

% \affiliation{%

% %%
% %% By default, the full list of authors will be used in the page
% %% headers. Often, this list is too long, and will overlap
% %% other information printed in the page headers. This command allows
% %% the author to define a more concise list
% %% of authors' names for this purpose.
% \end{comment}

%\title{Beyond Crawlers: Monitoring IPFS DHT Node Attributes using the Network Scale-up Method}

\author{Juan Marcos Ramirez}
\email{juan.ramirez@networks.imdea.org}
\author{Sergio Díaz-Aranda}
\email{sergio.diaz@networks.imdea.org}
\affiliation{%
  \institution{IMDEA Networks Institute}
  \city{Leganés}
  \state{Madrid}
  \country{Spain}
}

\author{Leonhard Balduf}
\affiliation{%
  \institution{TU Darmstadt}
  \city{Darmstadt}
  \country{Germany}}
\email{leonhard.balduf@kom.tu-darmstadt.de}

\author{Antonio Fernández Anta}
\affiliation{%
  \institution{IMDEA Software Institute \& IMDEA Networks Institute}
  \city{Madrid}
  \country{Spain}}
\email{antonio.fernandez@software.imdea.org}

%\title{An Alternative to Crawlers: Scalable P2P Network Attribute Tracking using the Network Scale-Up Method}
%\title{An Alternative to Crawlers: Scalable P2P Network Characterization using the Network Scale-Up Method}
\title{An Alternative to Crawlers: Scalable P2P Network Monitoring Using the Network Scale-Up Method}

%\title{Beyond Crawlers: Monitoring IPFS DHT Node Attributes using the Network Scale-up Method}

\begin{abstract}
\Ac{p2p} networks have remained an active area of research and development for decades. Originally used for data dissemination, they have later become instrumental to decentralized payment systems, blockchains, and Web3 applications following the introduction of Bitcoin by Satoshi Nakamoto in 2008.
Since decentralization is a crucial property of these networks,
they are continuously monitored to track attributes
such as the geographical distribution of nodes,
which serve as indicators of infrastructure concentration and resilience.
However, monitoring typically relies on full-network crawls
to derive complete, accurate snapshots,
which requires substantial computational and network resources.
% Our solution
In contrast, \ac{nsum}
is a statistical tool in social network analysis
that estimates the prevalence of subpopulations
based on responses to questions such as
\enquote{How many people do you know in a specific group?}
This work leverages the information embedded in Kademlia routing tables
to develop a lightweight \ac{nsum}-based measurement and estimation framework
that infers global network attribute distributions from a small set of nodes,
offering an alternative to full-network crawls.
As a case study,
we evaluate our approach on \ac{ipfs},
a well-researched and widely used Web3 data storage and retrieval network.
To the best of our knowledge,
%the proposed approach
this
is the first application of \ac{nsum} to decentralized network measurement.
% What we do
We evaluate our methodology on two datasets:
the first comprises 360 full-network crawls collected over one month
to assess our measurement-estimation approach across multiple simulated measurement scenarios,
while the second dataset was collected through a
%dedicated
deployment
of the proposed lightweight measurement framework during May 2026.
% Results
The results show that local Kademlia routing views, with only 10 nodes, preserve sufficient information
to accurately estimate global attribute distributions,
achieving normalized biases below \SI{10}{\percent},
and closely track temporal changes in network composition.

%These systems are often evaluated for their node populations and properties
%such as geolocation, attribution to hosting providers, or agent versions and protocol support,
%in order to characterize them and gauge their (de)centralization and resilience against provider or regional failures.
%These measurements rely on complete network crawls,
%with much research focusing on the speed and coverage of crawlers
%in order to derive accurate snapshots.
%We instead propose to use \ac{nsum},
%a statistical framework to estimate the prevalence of groups
%in low-resource scenarios using the neighborhood set of a small number of sampled nodes,
%to derive characterizations of \ac{p2p} networks.
%We compare resource usage and accuracy of our method
%using different sampling strategies
%against \hl{NUMBER} complete crawls of the \ac{ipfs} network
%in \hl{April} of 2026.
%We find that
%\ac{nsum} offers a less invasive alternative to existing crawling methods,
%using fewer resources,
%with a bias of less than \hl{4\%} in prevalence when sampling only \hl{1\%} of nodes.
%\lb{Can we maybe say that we need a constant number of nodes, which would make this infinitely scalable?}
\end{abstract}

\maketitle

% Reset acronyms from the abstract
\acresetall

%\lb{TODO for myself: Go through acronyms.}
%\lb{TODO for myself: non-breaking space before citations}
\lb{Target: CoNEXT, short paper, 10 pages + unlimited references + 2 pages appendix, deadline \hl{June 5th}}
%\lb{TODO: re-scope to P2P systems in general, use IPFS as a case study because well-researched and contemporary.}
\lb{tracking document for reviewer suggestions: https://www.overleaf.com/project/69ccf2c88eef6d6bc05086ad}
%\lb{TODO: All papers must include a statement or subsection about ethical issues raised by the work. (from the CfP)}
%\lb{Do we need the CCS classification?}
%\lb{I commented out the keywords, I think they're not mandatory and it saves us some space}

%%%%%%%%%%%%%%%%%%%%%%%%%%%%%%%%%%%%%%%%%%%
%%%%%%%%%%%%%%%%%%%%%%%%%%%%%%%%%%%%%%%%%%%
\section{Introduction}
%\lb{I'm wondering about motivation. Maybe we could claim that this method is less invasive than crawling all nodes? And faster?
%These are nice properties to measure P2P systems, because they always change due to churn, which necessitates fast crawling. Very large systems (BitTorrent) could then be measured in much shorter time with less resources. Using IPFS as an example is maybe nice because it's already one of the largest operational P2P networks at the moment, and it's a bit difficult to crawl due to the multiple addresses per peer etc., so it makes for a good system to validate the methods on?}
%\lb{Motivation: We could add something about IPFS being an example of a group of P2P systems, to which our techniques likely generalize. But this might trigger the question why we didn't do it on those systems... unsure. We could alternatively hint at measuring libp2p-based systems, of which IPFS is one instance. Other instances are Ethereum (the consensus layer, I think).}

The last two decades have seen a renewed interest in \ac{p2p} systems,
driven in part by the development of cryptocurrencies
such as Bitcoin and Ethereum~\cite{nakamoto2008bitcoin,wood2014ethereum,buterin2013ethereum,tschorsch2016bitcoin}
and distributed data storage systems such as \ac{ipfs}~\cite{benet2014ipfs},
forming the foundation of the larger Web3 movement to decentralize the Web.
%using technologies such as \ac{ens}~\cite{ens_foundation_ens_2017} and \acp{nft}~\cite{entriken_erc-721_2018} for identities and ownership
%and decentralized storage systems such as \ac{ipfs}~\cite{benet2014ipfs} for storage.
These modern \ac{p2p} systems are often measured for their properties,
such as size,
node populations,
or geospatial node distribution,
akin to measurements on earlier systems such as BitTorrent~\cite{cohen2003incentives,pouwelseBittorrentP2PFileSharing2005}
and Gnutella~\cite{DBLP:journals/corr/cs-DC-0209028}.
However, measuring \ac{p2p} systems at scale typically relies on active probing
and crawler-based techniques~\cite{Trudi2020Libp2p, Trudi2023Metric,DBLP:conf/imc/StutzbachR06}
which are inherently intrusive because they try to extract information
from \emph{all} reachable nodes
and pose challenges for capturing complete network snapshots due to churn~\cite{DBLP:conf/imc/StutzbachR06,stutzbach2006capturing}.
%and unresponsive nodes

%On the other hand, 
\Ac{nsum} is a statistical framework to estimate the proportion of
%hidden or hard-to-reach
subpopulations in social networks.
\ac{nsum} has been applied to diverse domains,
including disaster victims estimation~\cite{bernard1988many, bernard2001estimating},
and tracking diseases such as HIV/AIDS
and COVID-19~\cite{unaids2010guidelines, garcia2021estimating}.
Rather than relying on direct identification of nodes belonging to a specific sub-population,
\ac{nsum} uses \emph{indirect reporting}:
a sample of nodes is queried about how many of their neighbors belong to the target group.
\ac{nsum} offers two advantages:
it enhances privacy by avoiding direct disclosure,
and it increases coverage by enabling estimates even when full network visibility is unavailable.

\ac{ipfs}
implements a \ac{p2p} network
for decentralized content storage and retrieval~\cite{Ruizhe2024Closer}
aimed at the next generation of the Web.
It is built upon \lptp,
which forms the foundation of many modern \ac{p2p} systems,
such as Ethereum and Polkadot.
\ac{ipfs} is one of the largest \ac{p2p} networks in operation today~\cite{Trautwein2022Design}.
It has gained attention as an external storage layer for blockchain-based applications,
including \ac{nft} platforms,
and decentralized applications~\cite{Ascigil2019Towards, Chen2017Improved, Das2022Understanding, Hoang2020Privacy,balduf2022nft}.
\ac{ipfs} is particularly well-studied in the literature for its structure, content accessibility, and node distribution~\cite{Balduf2023Cloud, Ruizhe2024Closer,Lajam2021Performance, Balduf2022Monitoring, prunster2022total, Trautwein2022Design}.
The availability of tools to measure \ac{ipfs} accurately,
as well as the wide adoption of \lptp as an underlying technology for \ac{p2p} networks,
makes it a prime candidate to explore new measurement methodologies
to characterize modern \ac{p2p} systems.

\paragraph{Contributions}
This paper introduces a measurement and estimation framework for \ac{p2p} networks based on the \ac{nsum}.
By utilizing information in routing tables,
the proposed framework estimates the prevalence of node attributes
from partial network views,
offering an alternative to full-network crawls.
Our contributions are fourfold:
\begin{enumerate*}
\item We present, to the best of our knowledge, the first application of \ac{nsum} to decentralized network measurement and show how routing-table observations and node attributes can be mapped to the \ac{nsum} framework.
\item We design a lightweight measurement methodology that collects information from a small number of nodes and uses it with \ac{nsum} estimators to compute the prevalence of node attributes of the entire network.
\item We perform a large-scale evaluation using 360 IPFS \ac{dht} snapshots collected between April~1 and May~1, 2026, to assess the accuracy of the proposed framework under simulated measurement scenarios.
\item We implement the proposed lightweight measurement technique with 10 
\ac{ipfs} nodes to estimate the prevalence and track temporal changes in network attributes.
\end{enumerate*}
To evaluate the this framework, we consider geographic and cloud-provider distributions as representative attributes
and quantify estimation accuracy using prevalence estimates and normalized bias.

\paragraph{Results}
The routing table structure of the \ac{ipfs} Kademlia \ac{dht}
allows the implementation of a global estimator based on \ac{nsum}.
This method achieves high accuracy in estimating key network features,
such as the geospatial distribution of nodes and hosting patterns,
while requiring fewer resources than methods based on full-network crawls.
The estimation performance remains stable over time.
Through extensive simulations,
we evaluate
%the normalized bias of
two classical \ac{nsum} estimators
observing a \ac{nb} of less than \SI{10}{\percent}
when sampling just $10$ nodes.
Additionally,
our applied implementation of the method on $10$ real \ac{ipfs} nodes
shows that network features can be estimated accurately using only local information.
This approach is lightweight and faster than crawling,
allowing for more frequent snapshots,
which accurately track network features over time.
Our method offers a less invasive, faster, and resource-efficient approach to characterizing \ac{p2p} systems,
paving the way for more ethical measurement practices.
Compared to full-network crawls,
our method interacts with a fixed number of nodes
and finishes within seconds instead of minutes.
Our findings likely generalize to other \ac{p2p} networks with similar structures.

\section{Background}

\begin{comment}
This section introduces the framework for characterizing the IPFS network from a reduced number of samples using NSUM estimators.
We begin by formalizing IPFS as a directed graph and introducing key metrics such as node degrees and prevalence.
Then, we present the NSUM estimators and their theoretical foundations.
Finally, we explain how to implement the NSUM estimators to infer global network properties from limited observations.
\end{comment}

%%%%%%%%%%%%%%%%%%%%%%%%%%%%%%%%%%%%%%%%%%%
\subsection{The Network Scale-up Method}
\label{sec:background:nsum}

\ac{nsum} has been developed as a class of estimators
designed to compute an accurate estimate
of the size or prevalence of a subset of nodes%
---typically those sharing a specific attribute or property---%
using only the number of connections of a sample of nodes~\cite{Laga2021Thirty}.
Concretely,
the \ac{nsum} estimators are the estimators
that use only the number of outgoing connections that each sample node has
into several groups.
Given a network  $G = (V, E)$,
a subset of nodes $H$,
and sample nodes \(S \subset V\),
the \ac{nsum} estimators use two types of variables:
\begin{itemize}
    \item The out-degree,
    the total number of out-neighbor nodes of a node \( v \in V \) in the graph \( G \), denoted by \( R_v \).
    \item The number of out-neighbors of a node \( v \in V \)
    that belong to a given subset \( H \subset V \), denoted by \( C_{v,H} \).
\end{itemize}
The key intuition behind \ac{nsum} is
that the proportion of connections a given node \(v\) has to nodes in $H$,
expressed as \( C_{v,H} / R_v \),
carries meaningful information about the overall prevalence of that group in the network~\cite{Laga2021Thirty},
where the prevalence of $H$,
$\rho_H$,
is expressed as $\rho_H = \frac{|H|}{|V|}. $
% Building on this assumption,
% two classical \ac{nsum} estimators are commonly used.
% The first, known as the \textit{Mean of Ratios (MoR)}~\cite{killworth1998social},
% estimates prevalence as the average of the individual rates of the sample nodes $S$,
The representativeness of the ratios considers several random-mixing assumptions for the classical \ac{nsum} estimators. The first estimator, known as \textit{Mean of Ratios (MoR)}~\cite{killworth1998social}, assumes that $\mathbb{E}[C_{v,H} / R_v] = \rho_H$, and estimates prevalence as the average of the individual rates of the sample nodes $S$,
\[
\hat{\rho}^{MoR}_H = \frac{1}{|S|}\sum_{v \in S} \frac{C_{v,H}}{R_v}.
\]
The second estimator,
%referred to as the
\textit{Ratio of Sums (RoS)}~\cite{killworth1998estimation}, assumes that $C_{v,H}$ are independent variables and follow the Binomial model $C_{v,H} \sim Bin (R_v,\rho_H)$. The prevalence is calculated as the maximum likelihood estimator, which corresponds to the ratio between the sum of the out-neighbor nodes belonging to $H$ and the sum of the out-degree of all sampled nodes:
% The second,
% referred to as the \textit{Ratio of Sums (RoS)}~\cite{killworth1998estimation},
% is obtained as the maximum likelihood estimator
% of the Binomial model
% for the independent random variables $C_{v,H} \sim Bin (R_v,\rho_H)$.
% It estimates the prevalence
% as the ratio between the sum of the out-neighbor nodes belonging to $H$
% and the sum of the out-degree of all sampled nodes:
\[
\hat{\rho}^{RoS}_H = \frac{\sum_{v \in S} C_{v,H}}{\sum_{v \in S} R_v}.
\]
% \lb{reworded the following slightly, maybe check if still correct.}
A major advantage of the \ac{nsum} estimators is
that each sample node provides information about its neighbors.
As a result,
\ac{nsum} can leverage information from a substantially larger set of nodes
without requiring direct access to each one.
In particular,
studies in economic networks have shown
that this type of data reached the same conclusions
while reducing costs by \SIrange{70}{80}{\percent}~\cite{breza2020using}.
\subsection{Peer-to-Peer Networks and IPFS}
\label{sec:background:p2p}

\ac{p2p} networks are a class of overlay networks in which equal peers communicate to share resources
directly without the need for centralized servers.
Early applications of this technology focused on file sharing (\eg, BitTorrent~\cite{pouwelseBittorrentP2PFileSharing2005,cohen2003incentives}),
but the development of Bitcoin in 2008~\cite{nakamoto2008bitcoin} extended the concept
to cryptocurrencies.
\ac{p2p} networks can be divided into structured and unstructured networks
depending on their overlay structure and search procedures.
Unstructured networks such as Gnutella or Bitcoin use flooding mechanisms to disseminate messages through the network.
Structured networks such as \acp{dht} impose a logical structure upon the connections held by each node
and introduce a distance metric,
which ultimately allows efficient routing through the network towards a given target.
Each node in a structured network
maintains a routing table of other nodes.

\paragraph{Kademlia}
Kademlia~\cite{maymounkov2002kademlia} is a \ac{dht} protocol used in many modern \ac{p2p} systems (\eg, BitTorrent, \ac{ipfs}).
Each participant in Kademlia is identified by a pseudorandom ID in a large key space,
typically $[0,2^{256})$.
A node $N$ maintains connections in a routing table consisting of $k$-buckets,
 each of which holds up to $k$ entries.
The buckets are constructed based on the XOR distance $d = a \oplus b $ between the node ID of $N$ and the remotes.
The first bucket contains peers with which $N$ shares no common prefix,
with each consecutive bucket containing peers that share one more bit of common prefix.
%Thus, due to the limited size of the buckets,
%a node knows its \enquote{local neighborhood} in ID space exhaustively well,
%but further parts of the network are sampled.

\paragraph{Crawling}
It is possible to use the routing mechanism of structured overlays
to discover and enumerate the participants of the network
and their routing tables
through a process known as \emph{crawling}.
A crawler starts with a set of nodes
(\eg, the bootstrap nodes of a network),
and enumerates the contents of their routing table
through specifically targeted routing queries.
The newly learned nodes are added to a set of nodes to be crawled.
This process repeats until
%all crawling requests are finished and
no new peers are discovered.
For Kademlia-based \acp{dht},
a crawler typically issues
%a number of
requests targeted at specific $k$-buckets
of a node,
where each query returns up to $k$ new peers.
As typically only the first few buckets of a node's routing table are filled,
not all buckets need to be probed to get a complete snapshot of the network.
Numerous \ac{p2p} systems have been crawled
to describe their overlay structure and participants~\cite{Trudi2020Libp2p, Trudi2023Metric,DBLP:conf/imc/StutzbachR06}.

\paragraph{IPFS}
\ac{ipfs} is a contemporary \ac{p2p} network for data storage and retrieval~\cite{benet2014ipfs} built on \lptp.
It uses a Kademlia \ac{dht} for routing.
Nodes are identified via node IDs derived as the hash of their public key.
\lptp transparently multiplexes among multiple underlay connections,
meaning that nodes can advertise and be reachable at multiple addresses.
Underlay addresses in \lptp are encoded as multiaddresses~\cite{Trautwein2022Design}.
These multiaddresses support various IP versions (IPv4 or IPv6),
transport protocols (TCP or QUIC),
and network interfaces.
The node population is segregated into \ac{dht} servers and clients.
The former provide \ac{dht} functionality to other participants,
storing key-value mappings and answering queries,
whereas the latter only use the \ac{dht} without becoming a server themselves.
The network is particularly well-studied in the literature~\cite{Trautwein2022Design,Balduf2022Monitoring,Balduf2023Cloud,Ruizhe2024Closer}
for its participants, overlay structure, data stored, and performance.
A crawler for the \ac{ipfs} \ac{dht} exists~\cite{Henningsen2020Mapping}
that can enumerate all nodes and their routing tables.
The crawler uses crafted \textsc{find\_node} \acp{rpc} to retrieve specific $k$-buckets
of a remote node.
The crawler sends between $4$ and $20$ \acp{rpc} per node,
terminating once no new peers are learned.

\section{Methodology and Datasets}
\label{sec:data}

%%%%%%%%%%%%%%%%%%%%%%%%%%%%%%%%%%%%%%%%%%%
\subsection{NSUM IPFS Characterization}
\label{sec:data:nsum}

% This section presents an implementation of the \ac{nsum}
% to determine the size of a subset of nodes
% sharing a specific characteristic
% in a \ac{p2p} network,
% such as being located in a particular country
% or hosted by a specific cloud provider,
% using a reduced number of measurement nodes
% that are introduced in the network.
% We first describe the procedure for monitoring node attributes
% in the Kademlia \ac{dht}.
% Then,
% we analyze the factors that contribute to the effectiveness of the \ac{nsum}
% in this context
% and highlight its differences from the traditional applications
% of the methods in statistics and sociology.

We present an implementation of the \ac{nsum}
to estimate the prevalence of a subset of nodes
sharing a specific characteristic in the \ac{ipfs} \ac{dht},
such as being located in a particular country or hosted by a specific cloud provider,
using fewer measurement nodes.
In the following,
we describe our method
and then provide the theoretical justification behind the proposed \ac{nsum} implementation.

%Let $G_t=(V_t,E_t)$ a graph that represents the active nodes $V_t$ and the current connections $E_t$. 

Our method considers a set of measurement nodes $S$ in the \ac{ipfs} network.
We utilize information about their neighbors in the network,
specifically information of the peers in their buckets.
The nodes in $S$ can either be operated for explicit measurement purposes,
in which case their routing information can be obtained entirely locally,
or existing nodes in the \ac{ipfs} \ac{dht} can be crawled to obtain their buckets.
%A subset of nodes $S$ is introduced into the \ac{ipfs} \ac{dht},
%which collect information of their buckets over time.
%Recall (\cf~\Cref{sec:background:p2p}) how nodes in \ac{ipfs}
%are assigned uniformly random but stable node IDs at initial startup.
%These IDs determine in part the outconnections with the rest of the network
%through Kademlia's $k$-buckets.
We collect this routing information from $S$
at a point in time $t$.
%the nodes are refreshed to capture current connections on the DHT component,
%removing past connections with inactive nodes.
From metadata about the neighbors
(\eg, their underlay addresses),
we calculate the sought attributes of the peers in their routing table.
We can then measure the number of outconnections $R_{v,t}$,
and the number of out-neighbors $C_{v,H_t}$ that belong to $H_t$,
where $H_t$ is a set of nodes sharing one attribute.
We then apply the two estimators from \Cref{sec:background:nsum}
to arrive at the sought prevalence $\hat{\rho}_{H_t}$.

% \lb{The following paragraph feels weird, are the two formulas supposed to be the same? Are these really the reasons why the estimators are unbiased?}
Node attributes are assumed to be located uniformly at random in the \ac{ipfs} network from the perspective of the node ID space.
This is reasonable, as node IDs are assigned pseudorandomly,
not based on properties of 
the node.
%the underlay network.
In addition, peers are connected
%to active nodes
based solely on proximity in the node ID space.
For these reasons,
the random variables $C_{v,H_t}$ can be assumed to be independent and to follow the Binomial model $C_{v,H_t}\sim Bin(R_{v,t},\rho_{H_t})$,
which are the assumptions underlying the RoS estimator.
In particular, this implies that by the law of total expectation
\begin{multline}
   \mathbb{E} \left[ \frac{\sum_{v\in S} C_{v,H_t}}{\sum_{v\in S} R_{v,t}} \right]  = \mathbb{E} \left[ \mathbb{E} \left[ \frac{\sum_{v\in S} C_{v,H_t}}{\sum_{v\in S} R_{v,t}} \middle| R_v,v\in S \right]  \right] = \\ \mathbb{E} \left[ \frac{ \sum_{v\in S} \mathbb{E} \left[ C_{v,H_t}\right]}{\sum_{v\in S} R_{v,t}} \right] = \mathbb{E} \left[ \frac{ \sum_{v\in S} \rho_{H_t} R_{v,t} }{\sum_{v\in S} R_{v,t}} \right]  = \rho_{H_t}.  \nonumber
\end{multline}
Additionally, $\mathbb{E}[C_{v,H_t}/R_{v,t}] = \rho_{H_t}$, and therefore, MoR is unbiased because it is expressed as the sample mean of the variables $C_{v,H_t}/R_{v,t}$.

% Node attributes are assumed to be located uniformly at random
% in the \ac{ipfs} network from the perspective of the node ID space.
% This is reasonable,
% as node IDs are assigned pseudorandomly,
% not based on properties of the underlay network.
% Under this assumption,
% $\mathbb{E}[C_{v,H_t}/R_{v,t}] = \rho_{H_t}$,
% since the MoR is the sample mean of the variables $C_{v,H_t}/R_{v,t}$, the estimator MoR is unbiased.
% Furthermore,
% this implies that the variables $C_{v,H_t}$ are independent.
% This fact and the uniform distribution of node attributes
% imply that the variables $C_{v,H_t}$ follow the Binomial Model of RoS:
% \[
% C_{v,H_t}\sim Bin(R_{v,t},\rho_{H_t}).
% \]

% \sergio{This section presents an implementation of the NSUM to monitor a subset of nodes $H$ sharing a specific characteristic, such as being located in a particular country or hosted by a specific cloud provider, using a reduced number of measurement nodes $S$ that are introduced in the network. We show how to transform the data from the measures of the nodes of the Kademlia DHT to input data for NSUM estimation, and the reasons behind the proposal.}

% \sergio{Every node in the DHT component has access to the attributes of the peer in the routing table }

% \sergio{The connections in the DHT component of a Kademlia network depend only on the node IDs, which are assigned uniformly at random when a node accesses the DHT component} 

%%%%%%%%%%%%%%%%%%%%%%%%%%%%%%%%%%%%%%%%%%%
\subsection{Datasets and Experiments}
\label{sec:data:data}

In order to validate our method
and to showcase the feasible applicability in a real deployment,
we collect two main datasets:
\begin{enumerate*}
    \item A dataset consisting of full-network \emph{crawls}, and
    \item data from a deployment of ten \ac{ipfs} nodes,
    of which we collect the \emph{local bucket} information.
\end{enumerate*}
We then use these two datasets for two main experiments:
\begin{enumerate*}
    \item First,
    we utilize the crawls to simulate measurement studies with a number of randomly chosen measurement nodes $S$.
    This allows us to show the effectiveness of the NSUM estimators.
    %through the \ac{nb} as a function of sample size.
    \item Then,
    using the deployment of ten nodes,
    we use only their local routing information,
    \ie, the contents of their buckets,
    to drive the previously established estimators.
\end{enumerate*}
In both experiments,
we compare the results obtained through these estimations
to the ground truth of full-network crawls.

\begin{comment}
In this section,
we outline the data collection stage and provide details about the resulting datasets.
Specifically,
we collect two complementary datasets:
the first, referred to as the \textbf{IPFS-Benchmark Dataset},
was designed to assess the performance of the \ac{nsum} estimators in simulated measurement scenarios.
The second, referred to as the \textbf{Bucket Dataset},
incorporates the proposed measurement method for monitoring the attributes of the DHT IPFS network
using a reduced set of local nodes.
\end{comment}

\paragraph{Crawl Dataset and Simulated Measurement Studies}
To derive a ground truth for our estimators,
we crawl the entire \ac{ipfs} \ac{dht}
using an existing crawler~\cite{Henningsen2020Mapping,Trudi2020Libp2p}.
We crawl the network 12 times per day
between April~1 and May~1, 2026,
for a total of 360 snapshots.
The crawler enumerates all participants of the \ac{dht},
the edges between them,
and attributes of each node,
such as their underlay IP addresses,
client versions,
and supported protocols.
A crawl takes,
on average,
%WITH tmp1 AS (SELECT sample_round, MAX(end_timestamp) - MIN(start_timestamp) AS crawl_duration FROM 'crawled_nodes_2026-06-02.parquet' WHERE sample_round >= 137 AND sample_round <= 3009 GROUP BY ALL) SELECT MEAN(crawl_duration) FROM tmp1;
% > 00:06:48.464343
\SI{6}{\minute}~\SI{48}{\second},
with each single successfully crawled node taking on average
\SI{1.974}{\second}.
%WITH tmp1 AS (SELECT crawl_end_ts::TIMESTAMPTZ - crawl_begin_ts::TIMESTAMPTZ AS crawl_duration FROM 'crawled_nodes_2026-06-02.parquet' WHERE start_timestamp <= ('2026-05-01'::DATE + INTERVAL 2 HOURS) AND crawl_duration IS NOT NULL) SELECT MEAN(crawl_duration) FROM tmp1;
% > 00:00:01.973821
Each crawl generates two files:
\begin{enumerate*}
\item a metadata file with peer IDs, multiaddresses, and supported protocols, and 
\item a graph edge list showing directed P2P connections.
\end{enumerate*}
We remove nodes that are either unreachable or could not be fully enumerated.
%During our observation,
%\lb{To discuss maybe: I used the total number of nodes here, not the reachable/good ones, because it's higher. I think making IPFS look more relevant is good for us. On the other hand, the astute reader will notice that this does not fit together with the number of peers per bucket reported below.}
%\lb{keep 8k for now}
The network comprises \num{8166} nodes and \num{644747} edges on average.
% WITH tmp1 AS (SELECT sample_round, COUNT(*) AS num_nodes, COUNT(*) FILTER (WHERE connection_error IS NULL) AS num_good_nodes FROM 'crawled_nodes_2026-06-02.parquet' WHERE sample_round >= 137 AND sample_round <= 3009 GROUP BY ALL) SELECT MEAN(num_nodes) AS num_nodes, MEAN(num_good_nodes) AS num_good_nodes FROM tmp1;
%┌───────────────────┬───────────────────┐
%│     num_nodes     │  num_good_nodes   │
%│      double       │      double       │
%├───────────────────┼───────────────────┤
%│ 8165.647222222222 │ 4001.402777777778 │
%└───────────────────┴───────────────────┘
%WITH tmp1 AS (SELECT sample_round, COUNT(*) AS num_edges FROM 'crawled_graphs_2026-06-02.parquet' WHERE sample_round >= 137 AND sample_round <= 3009 GROUP BY ALL) SELECT MEAN(num_edges) AS num_edges FROM tmp1;
%┌───────────────────┐
%│     num_edges     │
%│      double       │
%├───────────────────┤
%│ 644747.0472222222 │
%└───────────────────┘

The objective of this dataset is to determine whether node attributes in the IPFS DHT can be accurately estimated
from a reduced set of sampled nodes,
thereby avoiding the need for exhaustive network crawls. 
For each snapshot,
we generate a measurement scenario by randomly selecting a subset of nodes from the network. 
Each sampled node exposes the peer IDs contained in its DHT routing table. 
Within the NSUM framework, these routing-table entries are treated as observable neighbors, while node attributes such as geographic location or cloud-provider affiliation define the target populations whose prevalence we aim to estimate.
Using only information collected from the sampled nodes, we estimate attribute prevalences through two classical NSUM estimators, MoR and RoS (\cf~\Cref{sec:background:nsum}).
To quantify estimation accuracy, we compute the normalized bias \ac{nb},
which measures the relative deviation between the estimated prevalence and the prevalence obtained from the corresponding crawl snapshot~\cite{diaz2026robust}:
\[
\mathrm{NB} = \frac{1}{N}\sum_{i=1}^N \frac{|\hat{\rho}_i - \rho|}{\rho},
\]
where \(N\) is the number of realizations, \(\rho\) is the ground truth, and \(\hat{\rho}_i\) is the \ac{nsum} prevalence estimate in the $i$-th realization.

In terms of complexity when compared to complete network crawls,
this scenario considers crawls of $|S|$ randomly chosen nodes,
and thus utilizes approximately a fraction $|S| / |V|$ of \acp{rpc} with respect to a complete crawl of all $V$ nodes of the \ac{dht}.
On the other hand,
if we were to operate any node in $S$ for measurement purposes specifically,
its bucket contents would be available locally,
thus eliminating the need to crawl it.
Running the \ac{nsum} estimators themselves takes less than \SI{1}{\milli\second}.
As such,
the application of \ac{nsum} is much faster than crawling,
and reduces the number of \acp{rpc} issued for measurement purposes.
The former is important as
previous works have shown that \ac{p2p} networks
must be measured quickly due to churn~\cite{stutzbach2006capturing,DBLP:conf/imc/StutzbachR06,falkner2007profiling}.

\begin{comment}
To analyze the performance of the NSUM estimators in simulated measurement scenarios, we collected multiple snapshots of the IPFS DHT network using the crawler described in~\cite{Henningsen2020Mapping, Trudi2020Libp2p}. Specifically, the crawler identifies reachable nodes within the libp2p DHT/Kademlia overlay and reconstructs their neighborhood graph. For each discovered peer, the crawler records the peer identifier, all multiaddresses obtained from the DHT, routing table entries, and supported protocols. The crawler uses multiple \texttt{FindNode} requests to target specific DHT buckets, enabling accurate recovery of routing table information. It starts with a set of bootstrap nodes and iteratively queries routing tables, connecting to new peers until all reachable nodes are discovered.
%This approach supports large-scale reconstruction of the DHT topology and facilitates analysis of connectivity patterns, peer distribution, and network-level properties of IPFS/libp2p overlays. 
To collect this dataset, we conducted 12 crawls daily from April 1 to May 1, 2026, resulting in 360 snapshots.
\end{comment}

\paragraph{Bucket Dataset and Real-World Deployment}
In addition to the above complete crawls,
we also operate ten \ac{ipfs} nodes
spread among three servers in Germany,
and collect their local routing tables.
This experiment shows the practical applicability of our proposed method
and measures the accuracy of the derived results.
Five nodes operate as \ac{dht} servers,
and their routing tables are already part of the above crawl dataset.
%but furthermore allow us to verify the correctness of the crawls.
The other five nodes operate as \ac{dht} clients,
which
%makes them invisible to crawling,
%as they are never included in other nodes' routing tables,
%but 
allows them to operate with fewer resources.
\Ac{dht} clients still maintain a routing table in order to \emph{use} the \ac{dht}.
Using the routing tables of these ten nodes,
we can evaluate whether it is possible to utilize the \ac{nsum} estimators
from a purely local perspective.
%\lb{check if true once results are in}
%operating as few as one node.
For each node,
we extract 
its routing table every \SI{15}{\minute}
between May~26 and June~1, 2026,
resulting in 596 snapshots,
while continuing to collect crawls as described above.
Each node maintains an average of \num{169} peers in its buckets.
%WITH tmp1 AS (SELECT host, node, sample_round, COUNT(*) AS num_entries FROM 'buckets_2026-06-02.parquet' WHERE sample_round >= 5442 AND sample_round <= 6037 GROUP BY ALL) SELECT  MEAN(num_entries) FROM tmp1 ;
%┌───────────────────┐
%│ mean(num_entries) │
%│      double       │
%├───────────────────┤
%│ 169.4557046979866 │
%└───────────────────┘

%\lb{TODO move to results or discussion maybe, depending on results.}
Kademlia routing tables are affected by temporal dynamics.
In particular,
as a security mechanism,
bucket management policies prioritize long-lived,
stable peers,
preferentially retaining recently responsive nodes and evicting transient or unstable ones.
Consequently,
routing tables
of long-lived nodes
tend to over-represent highly available peers and under-represent churn-prone nodes.
%\lb{Does this not also apply to our benchmark dataset then? Because those buckets should also prefer long-lived nodes.} \jm{You are right. Reviewers may think that. However, the benchmark dataset's good performance stems from the random selection (sampling) of nodes, which chooses nodes with different connection times. Seems that the sampling process selects a large proportion of ``fresh'' nodes. In our case, churn is an advantage because nodes have different connection times and node lists with different survivor times.}
%\lb{hmm hmm. Let's hope they don't notice :)}
This effect may introduce survivorship bias into the observed bucket lists,
increasing the probability of overlap among sampled nodes for stable peers.
To mitigate this potential bias,
we restart each node after every measurement round,
which causes it to clear its routing table and re-join the network.
%This procedure avoids artifacts introduced by the Kademlia protocol's bucket eviction and peer-retention mechanisms,
%ensuring that each snapshot reflects a newly reconstructed neighborhood view of the network.

\paragraph{Feature Assignment}
\begin{comment}
From the underlay IP address
we derive features that are typically of interest
in \ac{p2p} characterizations:
We derive the geolocation using \hl{TODO}
and attribute the address to a cloud provider and ASN using \hl{TODO}.
Both methods operate offline
and address information never leaves our machines.
In case of multiple underlay addresses,
we take a majority vote to set the feature value for the node.
\end{comment}
Recall (\cf~\Cref{sec:background:p2p}) how each node in the IPFS DHT network
is uniquely identified by a peer ID.
%derived from the hash of its public key.
Metadata per node includes multiaddresses,
which
%are lists of IP addresses that
identify the network endpoints to reach the node.
We map each IP address
%in the multiaddress list
of a node to a geolocation
using the MaxMind GeoLite2 database~\cite{GeoLiteIP}
and the Udger database~\cite{udgerDB}
to link the IP addresses to known cloud providers.
Because a single peer may advertise multiple IP addresses,
the inferred attributes may not be consistent across all endpoints.
For example,
different addresses of the same node may resolve to distinct geographic locations or cloud infrastructures
due to multi-homing,
VPN usage,
or distributed cloud deployments.
To assign a 
single
%consistent representative
attribute to each peer ID,
we adopt the majority-vote strategy proposed by Balduf~\etal~\cite{Balduf2023Cloud}.
This approach selects the most frequently observed geolocation or cloud provider for that node based on resolved IP addresses.

\section{Results}
\label{sec:results}

%%%%%%%%%%%%%%%%%%%%%%%%%%%%%%%%%%%%%%%%%%%
\subsection{Estimator Benchmark}
\label{sec:results:benchmark}

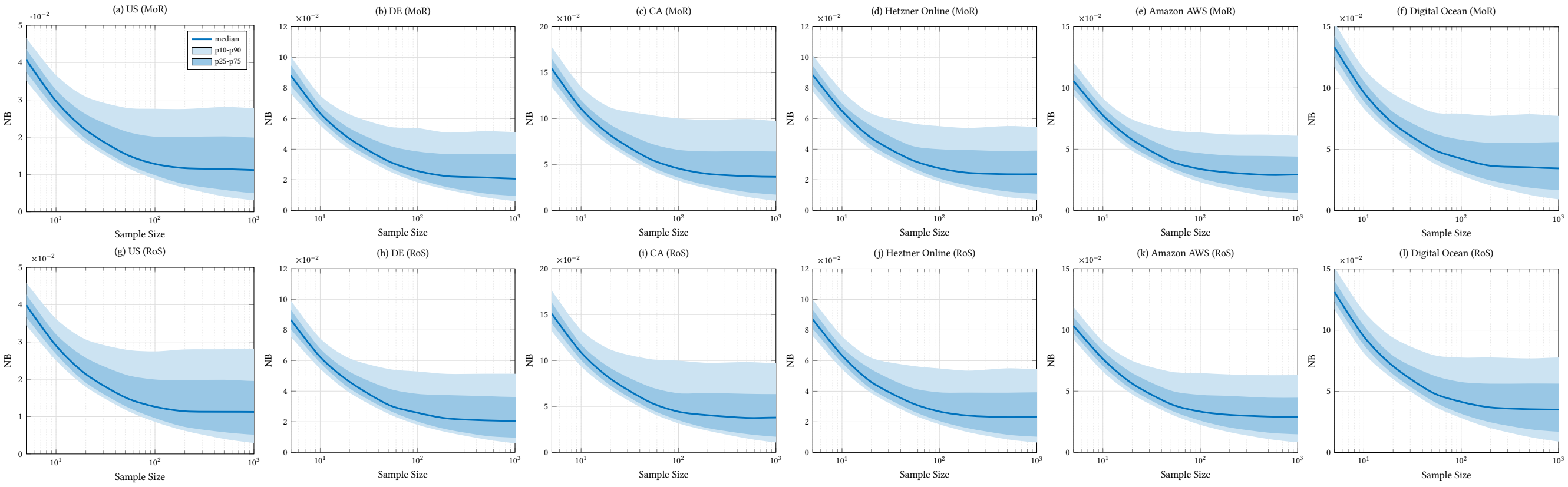
\begin{figure*}
\begin{center}
    \input{figtex/nb_plots}
\end{center}
    \caption{IPFS-Benchmark Dataset. Normalized bias (NB) of prevalence estimates as a function of sample size for six IPFS node attributes. The top row shows results are obtained with the MoR estimator, and the bottom row with the RoS estimator. Solid lines represent the median NB across the 360 snapshots, whereas the shaded regions indicate the $p25$-$p75$ and $p10$-$p90$ percentile intervals.}\label{fig:nb_vs_size}
\end{figure*}

We begin by assessing the proposed \ac{nsum}-based measurement framework
using the 360 snapshots of the IPFS-Benchmark Dataset
as described in \Cref{sec:data:data}.
%Our objective is to determine whether node attributes in the IPFS DHT can be accurately monitored from a reduced set of sampled nodes, thereby avoiding the need for exhaustive network crawls. For each snapshot, a measurement scenario is generated by randomly selecting a subset of nodes from the network. Each sampled node exposes the peer IDs in its DHT routing table. In the \ac{nsum} analogy, these routing-table entries are treated as observable neighbors, and each peer is associated with one or more attributes, such as its country or cloud provider. In \ac{nsum}, these attributes represent the target population for which we aim to estimate prevalence. Using only information from the sampled nodes, we estimate the prevalence of the target attributes with two classical \ac{nsum} estimators: the Mean of Ratios (MoR) and the Ratio of Sums (RoS). For a given sample size, the experiment is repeated 100 times, each time selecting a new random subset of nodes. 
%The resulting estimates are compared with the ground-truth prevalence computed from the corresponding crawl snapshot. 
Fig.~\ref{fig:nb_vs_size} displays the median \ac{nb}
together with the $p10$-$p90$ and $p25$-$p75$ percentile intervals
as a function of the sample size.
Results are shown for six representative attributes:
three countries (the United States (US), Germany (DE), and Canada (CA))
and three cloud providers (Hetzner Online, Amazon AWS, and Digital Ocean).
The curves correspond to aggregates across the 360 snapshots.
% The NB quantifies the relative deviation of the estimated prevalence from the ground truth~\cite{diaz2026robust}, averaged over multiple realizations. This metric is defined as
% \[
% \mathrm{NB} = \frac{1}{N}\sum_{i=1}^N \frac{|\hat{\rho}_i - \rho|}{\rho}
% \]
% where \(N\) is the number of realizations, \(\rho\) is the ground truth, and \(\hat{\rho}_i\) is the \ac{nsum} prevalence estimate in the $i$th realization. 

%As shown in Fig.~\ref{fig:nb_vs_size},
The NB decreases monotonically with sample size
across all attributes and both estimators.
This behavior is expected
because larger samples provide broader coverage of the \ac{dht} routing state
and reduce sampling variability,
yielding more accurate prevalence estimates.
The largest gains occur when the sample size increases
from about 5 to 50 nodes.
Beyond roughly 100 nodes,
the curves flatten,
indicating diminishing returns.
Secondly,
even for sample sizes below 20 nodes,
the \ac{nb} remains relatively low.
For instance,
for attributes such as US, DE, and Hetzner Online,
the median NB rapidly falls below \SI{5}{\percent}
with fewer than 50 sampled nodes.
This result is particularly relevant
because it suggests that accurate monitoring can be achieved
using only a very small fraction of the network.
Lastly,
attributes with low prevalence are more difficult to estimate.
In particular, CA and Digital Ocean consistently show errors larger than those in the US or Hetzner Online.
This difference arises because \ac{nsum} estimators depend on observing a sufficient number of neighbors
that share the target attribute.
Consequently,
for low-prevalence attributes,
fewer observations are available,
leading to greater variance and larger errors.
Finally,
RoS and MoR show very similar behavior,
with both exhibiting nearly identical trends.
This suggests that the main source of uncertainty is sampling variability,
not estimator choice.
RoS has a slight advantage at larger sample sizes,
but the difference is small.

\begin{framed}
\textbf{Takeaway 1:}
%Our results show that
\Ac{nsum}-based estimation of \ac{ipfs} attributes
is feasible using a small number of sampled nodes.
For all evaluated countries and cloud providers,
both RoS and MoR achieve low \ac{nb}
with sample sizes well below those of full-network crawls.
These findings indicate that the local views maintained by Kademlia routing tables
preserve sufficient information about global attribute distributions
to enable accurate estimation from partial observations and classical \ac{nsum} estimators.
\end{framed}

\begin{figure*}
\begin{center}
    \input{figtex/bckt_results}
\end{center}
    \caption{IPFS-Benchmark Dataset. Comparison of crawl-derived attribute prevalences with \ac{nsum} estimates (MoR and RoS) from 10 randomly sampled nodes at a snapshot captured on April 5, 2026. (a) country and (b) cloud-provider. (c)–(h) Temporal evolution of selected attributes. Evolution of (i) Non-cloud and (j) Cloud.
    % \lb{change to Full-crawl, as in Figure 3}
    }
    \label{fig:temporal}
\end{figure*}
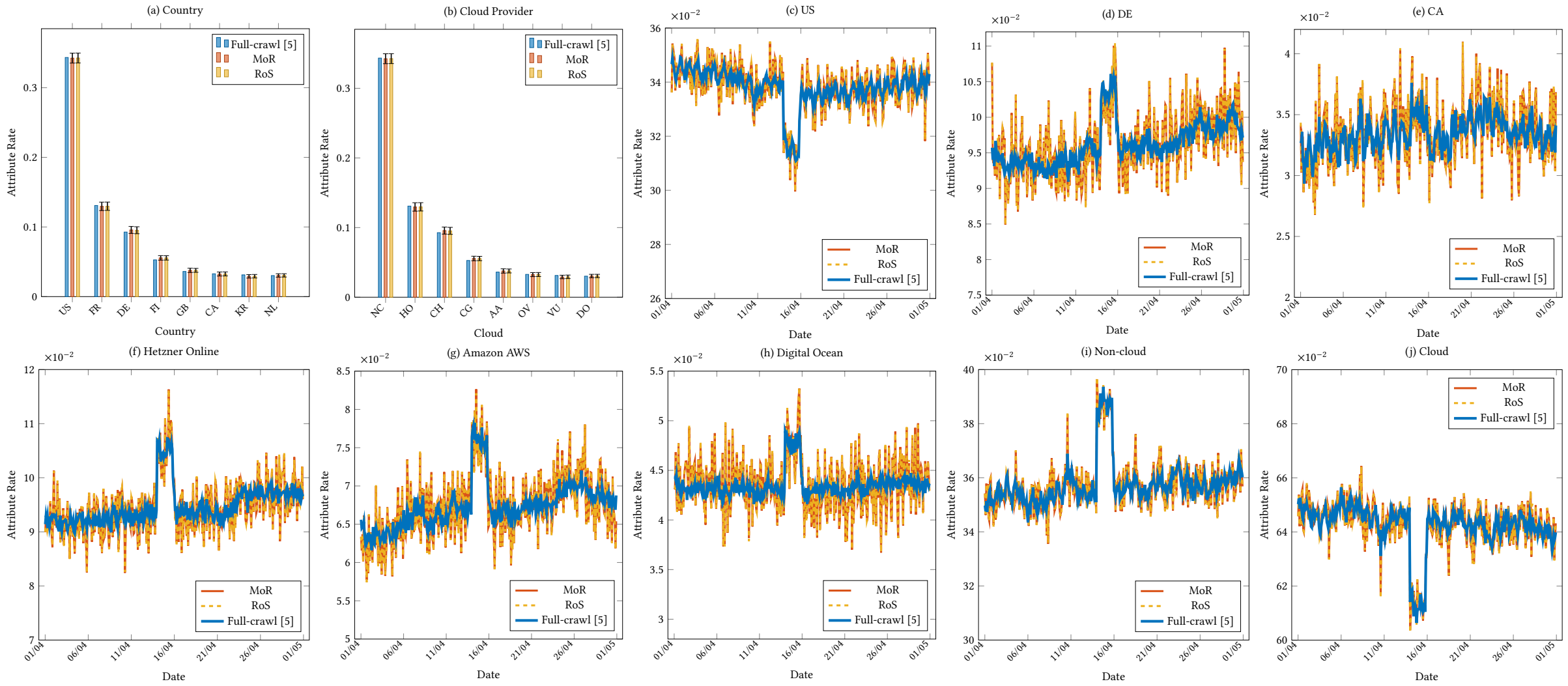

Figures~\ref{fig:temporal}(a) and \ref{fig:temporal}(b) compare the attribute prevalence estimates obtained using the MoR and RoS estimators against the prevalence values derived from a complete IPFS DHT crawl following the method of Balduf \textit{et al.}~\cite{Balduf2023Cloud}. The results correspond to a single snapshot collected on April 5, 2026, at 16:17 UTC.
To simulate a real measurement scenario with limited server availability,
the procedure randomly selects only 10 nodes from the network.
For each estimator,
the experiment is repeated 100 times,
with a different set of sampled nodes in each realization.
The reported MoR and RoS values correspond to the average prevalence across realizations,
and the error bars represent the corresponding \SI{95}{\percent} empirical confidence intervals.
For visualization purposes,
cloud-provider names are abbreviated as follows:
% \lb{Unknown sounds suspicious, is this Non-Cloud?}
% \lb{TODO change to non-cloud}
NC (Non Cloud), HO (Hetzner Online), DO (DigitalOcean), VU (Vultr), AA (Amazon AWS), CH (Choopa), CG (Contabo GmbH), and OV (OVH).

For both countries and cloud providers,
the estimates from MoR and RoS closely match the prevalence values from the crawl.
Although the corresponding confidence intervals are slightly wider for attributes with lower prevalence,
such as CH, CA, and DO,
the estimated attribute rates remain close to the crawl-derived values.
In addition, the MoR and RoS estimates are nearly indistinguishable across all attributes.
This observation is consistent with the previous results shown in \Cref{fig:nb_vs_size}, where both estimators exhibited similar NB values.

Figures~\ref{fig:temporal}(c)--\ref{fig:temporal}(h) extend this analysis to the temporal domain.
The plots show the evolution of prevalence estimates 
across three countries (US, DE, and CA)
and three cloud providers (Hetzner Online, Amazon AWS, and Digital Ocean)
based on
%only
10 sampled nodes.
The ground-truth time series are derived from complete network crawls.
Finally,
Figures~\ref{fig:temporal}(i) and \ref{fig:temporal}(j)
present the temporal evolution of the aggregate Cloud and Non-cloud populations estimated using MoR and RoS,
respectively.
In all cases,
the \ac{nsum} estimates follow the temporal dynamics observed in the crawl-derived ground truth.
While the \ac{nsum} estimates exhibit greater short-term variability,
they accurately capture major changes in network composition,
including transient decreases in the prevalence of US nodes and corresponding adjustments in the distribution of cloud providers.
These results indicate that a small number of sampled routing tables is sufficient
not only to estimate attribute prevalence,
but also to monitor temporal variations in the structure of the IPFS DHT.

\begin{framed}
\textbf{Takeaway 2:}
The temporal dynamics of the IPFS DHT can be accurately monitored using a small number of randomly selected nodes.
Even with data from just 10 nodes' routing tables,
the MoR and RoS estimators accurately reflect both the current prevalence
and the temporal evolution of attributes observed in complete network crawls.
%The random selection of nodes is a crucial aspect of our simulated measurement method,
%as it helps prevent temporal biases arising from bucket management policies that favor long-lived peers.
%\lb{does this not assume that we get a lot of \enquote{fresh} nodes in this selection of 10 nodes? Because any long-lived node we select should have the same biases, no?}
%\lb{maybe we can frame this is a plus for our method, because churn is good for us, but bad for everyone else. But it's difficult to sell the idea of randomly sampling nodes in the DHT \emph{wihout} crawling the entire DHT first, so I'd leave that for future work :)}
%\lb{hmm, we're selling the random sampling here, but that's difficult without a complete crawl. You have to start crawling somewhere, \ie, the bootstrap nodes (extremely long-lived), and then try to find random nodes from there, without crawling everyone... difficult.}
\end{framed}

\begin{figure*}
\begin{center}
    \input{figtex/sim_results}
\end{center}
    \caption{Bucket Dataset. Comparison of crawl-derived attribute prevalences with \ac{nsum} estimates (MoR and RoS) using the proposed measurement technique with 10 local nodes at a single snapshot. (a) country (b) cloud-provider. (c)–(h) Temporal evolution of selected attributes.  Evolution of (i) Non-cloud and (j) Cloud.}
    %\lb{(i) and (j) are missing the ground-truth data}
    \label{fig:buckets}
\end{figure*}
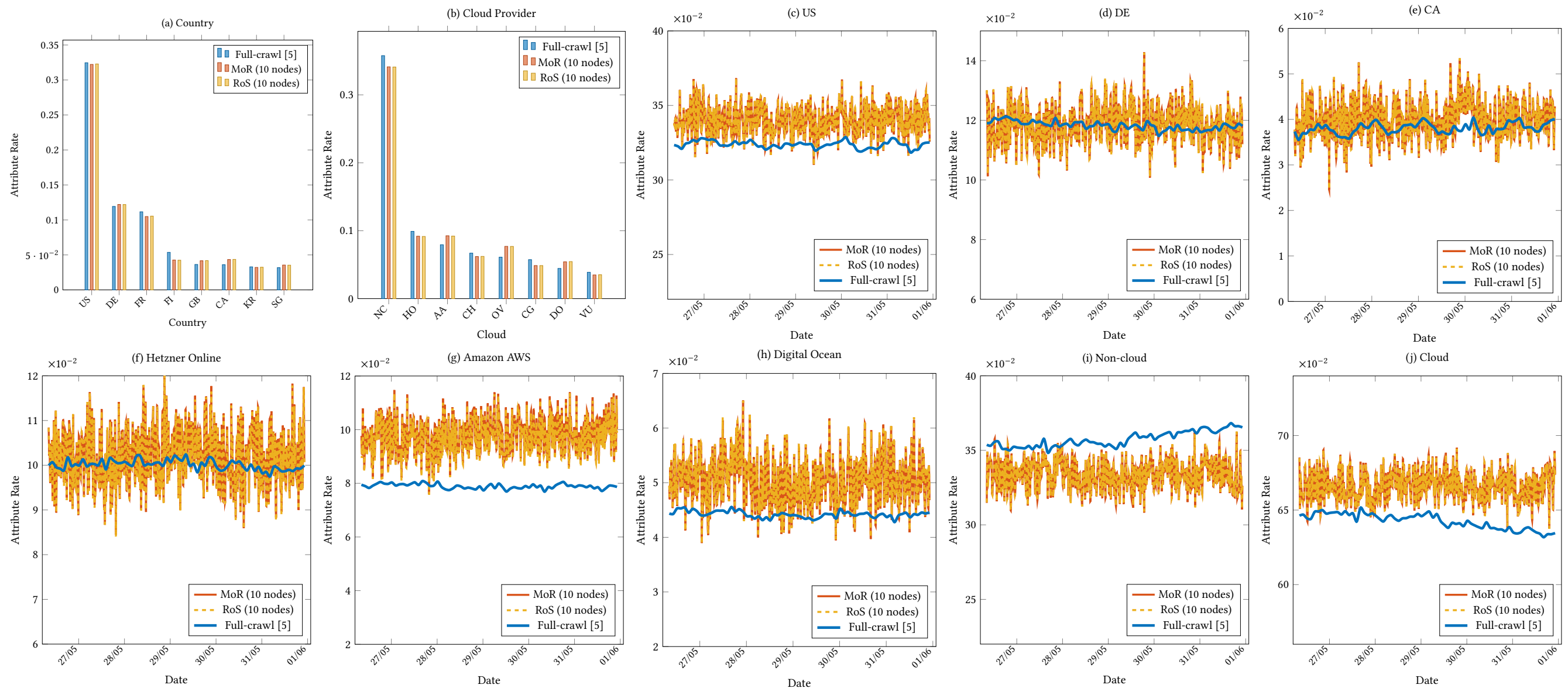

%%%%%%%%%%%%%%%%%%%%%%%%%%%%%%%%%%%%%%%%%%%
\subsection{Implementation Results}
\label{sec:results:implementation}

Figures~\ref{fig:buckets}(a) and \ref{fig:buckets}(b) compare the attribute prevalences from a full-network crawl
with estimates from the MoR and RoS estimators,
using the proposed lightweight measurement framework (\cf~\Cref{sec:data:data}).
The results are based on a single snapshot
and cover two classes of node attributes:
countries and cloud providers.
\ac{nsum} estimates are derived from information collected locally from 10 nodes.
%As shown in the figure,
Both MoR and RoS closely match the attribute distributions from the full crawl
across all evaluated countries and cloud providers.
The agreement is particularly notable given the small number of sampled nodes.
Although minor deviations can be observed in attributes with lower prevalence,
the overall ranking and relative proportions of the categories are accurately preserved.
From this static perspective,
the results suggest that even a limited number of local Kademlia routing table observations
contain sufficient information to recover global attribute distributions.

Figures~\ref{fig:buckets}(c)--\ref{fig:buckets}(h) extend the analysis to the temporal domain
by comparing crawl-derived prevalences
with estimates from the proposed measurement and estimation framework
for three countries (US, DE, and CA)
and three cloud providers (Hetzner Online, Amazon AWS, and Digital Ocean).
The results are based on one week of measurements
collected between May 26 and June 1, 2026.
Owing to its significantly lower measurement overhead,
the proposed framework enables more frequent observations:
in this case,
we collect routing-table snapshots from 10 local nodes every \SI{15}{\minute}
%(four snapshots per hour),
whereas full-network crawls are performed once every two hours.
%While it is possible to crawl the \ac{dht} more frequently,
%each crawl creates load on the network,
%in contrast to our proposed method.
Although the \ac{nsum} estimates
exhibit greater variability
than the crawl-derived measurements,
they consistently track the temporal evolution of the monitored attributes.
MoR and RoS estimates capture trends in network composition,
despite relying on a very limited number of sampled nodes.
This behavior indicates that lightweight routing-table measurements can be used
not only to estimate attribute prevalence
but also to monitor temporal changes in the IPFS DHT. Appendix \ref{sec:app} presents the estimation results for 5 nodes with NAT and 5 nodes without NAT.

\begin{framed}
\textbf{Takeaway 3:}
Despite observing only 10 sampled nodes,
the proposed NSUM-based measurement and estimation framework
accurately captures both the static distributions
and the temporal evolution of attributes in the IPFS DHT.
These results indicate that local Kademlia routing-table views
preserve sufficient information about the network composition,
enabling monitoring and temporal characterization
without repeated full-network crawls.
\end{framed}

\section{Discussion}
\label{sec:discussion}

We were able to show that \ac{nsum} is applicable to \ac{p2p} measurements.
This implementation differs from traditional applications of \ac{nsum} in statistics and sociology,
and characteristics of \ac{ipfs} actually mitigate limitations commonly encountered in those domains.

In the traditional application of \ac{nsum} in social networks,
a representative sample of the population is required.
Representative sampling is necessary in social networks
because individuals often have different propensities
for forming connections,
and neighbors in the sample report the common contacts,
particularly in highly clustered networks.
These issues are largely absent in
%\ac{ipfs},
\ac{p2p} networks,
where
%nodes and
attributes are uniformly distributed in the ID space.
Another challenge found in social networks is the reporting behavior.
Information is provided by sampled individuals
and may be affected by recall errors,
malicious responses,
and a lack of transmission from the contacts~\cite{Laga2021Thirty}.
In contrast,
peer information in IPFS is obtained directly from routing tables,
eliminating these sources of reporting error
and providing accurate measurements of node attributes of the peers.
As for the complexity of our proposed method,
we acknowledge that,
while it greatly reduces the number of participants contacted
and thus load on the network being generated,
it cannot eliminate them entirely.
Even in our proposed implementation using local bucket information
%of a number of measurement nodes,
the nodes themselves need to interact
with the network prior to our measurements.
%particularly given the security enhancements of the \lptp Kademlia implementation.
Lastly,
our simulations use randomly sampled nodes,
and our implementation uses ten dedicated nodes.
A middle ground and a potential solution to the question of how to randomly sample
a number of nodes in a DHT,
could be to operate just one measurement node and randomly sample among its peers.

%%%%%%%%%%%%%%%%%%%%%%%%%%%%%%%%%%%%%%%%%%%
%%%%%%%%%%%%%%%%%%%%%%%%%%%%%%%%%%%%%%%%%%%
\section{Related Work}

The popularity of IPFS has led to increased research focused on characterizing and understanding it.
These studies have provided valuable information on the architecture of IPFS, including its P2P protocol design and node distribution.
For instance, Trautwein et al.~\cite{Trautwein2022Design} performed a large-scale measurement study that revealed the global reach of IPFS nodes across more than 2,700 Autonomous Systems and 152 countries, emphasizing its potential for decentralized content delivery.
Other efforts have utilized crawler-based methods~\cite{Trudi2023Metric} to examine the topology and dynamics of IPFS.
Henningsen et al.~\cite{Henningsen2020Mapping} analyzed connectivity and churn.
More recently, Balduf et al.~\cite{Balduf2023Cloud} conducted
a multimodal measurement study to examine the topology,
traffic patterns,
and hosting environments.
These approaches rely on extensive crawling and probing, which can be resource-intensive and difficult to scale in decentralized environments.
A number of works have proposed sampling techniques to accelerate estimates of the size or number of items in large-scale \ac{p2p} systems such as KAD and the BitTorrent \ac{dht}~\cite{DBLP:conf/p2p/WangK13,steiner2007global,steiner2007actively,DBLP:conf/iptps/MemonRGS09}.
These usually depend on crawling part of the entire keyspace exhaustively,
which still potentially generates considerable load in the network.

In contrast,
our approach introduces a lightweight, statistically grounded framework
to characterize the IPFS network via \ac{nsum}
using partial observations and indirect reporting.
%The shares of participants in specific countries presented in earlier works differs from our numbers,
%but this is due to changes in the network,
%as we utilize the same tools and methods as the referenced works to derive our ground truth dataset.

%Other works have explored the trade-offs between decentralization and performance, such as the integration of centralized components like HTTP gateways and DHT boosters to improve content retrieval latency. Further investigations have examined the evolving centralization trends within IPFS, showing that a small fraction of nodes—often hosted in cloud environments—serve a disproportionate share of content. Collectively, these efforts underscore the complexity of maintaining decentralization while ensuring performance and scalability in large-scale P2P systems.

%Other studies have investigated the trade-offs between decentralization and performance, highlighting the role of centralized components such as HTTP gateways and DHT boosters in improving content availability and retrieval latency~\cite{Wei2024Tussle}. Additional research has analyzed the structure and dynamics of IPFS gateways, uncovering usage patterns, content popularity, and the impact of node churn on system stability~\cite{Xu2023Dissecting}. Collectively, these efforts provide a comprehensive view of IPFS as a complex, evolving ecosystem that balances decentralization with practical performance considerations.

%%%%%%%%%%%%%%%%%%%%%%%%%%%%%%%%%%%%%%%%%%%
%%%%%%%%%%%%%%%%%%%%%%%%%%%%%%%%%%%%%%%%%%%
\section{Conclusions and Future Work}

The IPFS DHT protocol enables the estimation of feature prevalence using a small sample set.
%\lb{careful with \enquote{balanced} --- the prefix tree of Kademlia is not balanced.}
%\lb{What does \enquote{efficient characterization} mean here? Who is characterizing what, and why?}
Building on this property, classical \ac{nsum} estimators (RoS and MoR) provide accurate and consistent prevalence estimates of global properties, such as geographic distribution and hosting patterns,
even when applied to random samples as small as 10 nodes.
This finding underscores the potential of \ac{nsum} for low-overhead and less intrusive monitoring of large decentralized networks.
%This finding underscores the potential of NSUM for measuring attributes of IPFS with low resources and less intrusive methods, enabling faster and more ethical procedures.
% \lb{I'd be careful with \enquote{monitoring}, as this always holds a temporal connotation for me. I'd maybe just focus on \enquote{measuring}, with low resources, non-intrusive, potentially ethically better, \etc}
While low-prevalence groups require slightly larger samples for precise estimation, the overall bias remains manageable, validating \ac{nsum}'s robustness under realistic constraints.

Looking ahead, we plan to extend this approach beyond DHT nodes to incorporate all network participants.
Furthermore, we aim to explore its applicability to other decentralized networks, such as Ethereum or other networks for which complete enumeration is difficult, and investigate adaptive sampling strategies that further reduce measurement overhead while maintaining accuracy.
Finally,
a current limitation of our work is that,
while we can estimate the prevalence of certain subgroups,
a measurement of the network size is needed to calculate the number of nodes per group.
%, which we plan to implement in the future.
%\lb{Filecoin is less relevant these days, I feel like. We could claim to extend this to Ethereum, or networks for which complete enumerations are difficult (Bitcoin, maybe?).}

\appendix

\section*{Ethics and use of Generative AI}

Our study raises a few ethical concerns.
The crawls and routing tables we collect include IP addresses, but we do not attempt to associate these with the identities of specific individuals. 
%For the collected data, we replace IP addresses by their derived features ensure a certain level of anonymity.
For geolocation and cloud-hosting, we use the MaxMind GeoLite2 and Udger databases~\cite{GeoLiteIP,udgerDB},
which operate locally on our machines.
Our work focuses solely on estimating network characteristics, not on behavior or personal data.
%We do not perform individual analyses,
%nor do we 
We do not collect any additional datasets beyond those described in the article.
In particular, our research does not involve handling or analyzing content data.
As for generative AI,
we use ChatGPT to revise previously written text.
No text was authored solely by generative AI.

\section*{Acknowledgments}

The work of Sergio Díaz-Aranda has been funded by \textit{Comunidad de Madrid} predoctoral grant PIPF-2022/COM-24467.
This paper has been funded by project PID2022-140560OB-I00 (DRONAC) funded by MICIU/AEI /10.13039/501100011033 and ERDF, EU.
This research is part of the Grant Excelencia María de Maeztu CEX2024-001471-M funded by MICIU/AEI/10.13039/501100011033.

%%%%%%%%%%%%%%%%%%%%%%%%%%%%%%%%%%%%%%%%%%%%%%%%%%%%%%%%%%%%%%%%%%%%%%%%%%%%
% We're in the endgame now

\bibliographystyle{ACM-Reference-Format}
\bibliography{refs}

% \onecolumn

\clearpage

\appendix

\setcounter{equation}{0}
\setcounter{figure}{0}
\setcounter{table}{0}
\renewcommand{\theequation}{A\arabic{equation}}
\renewcommand{\thetable}{A\arabic{table}}
\renewcommand{\thefigure}{A\arabic{figure}}

\begin{figure*}[t!]
\begin{center}
    \input{figtex/nonat_results}
\end{center}
    \caption{Bucket Dataset. Comparison of crawl-derived attribute prevalences with \ac{nsum} estimates (MoR and RoS) using the proposed measurement technique with 5 NATed nodes at a single snapshot. (a) country (b) cloud-provider. (c)–(h) Temporal evolution of selected attributes.}\label{fig:nat}
\end{figure*}

\begin{figure*}
\begin{center}
    \input{figtex/nat_results}
\end{center}
    \caption{Bucket Dataset. Comparison of crawl-derived attribute prevalences with \ac{nsum} estimates (MoR and RoS) using the proposed measurement technique with 5 NoNATed nodes at a single snapshot. (a) country (b) cloud-provider. (c)–(h) Temporal evolution of selected attributes.}\label{fig:nonat}
\end{figure*}
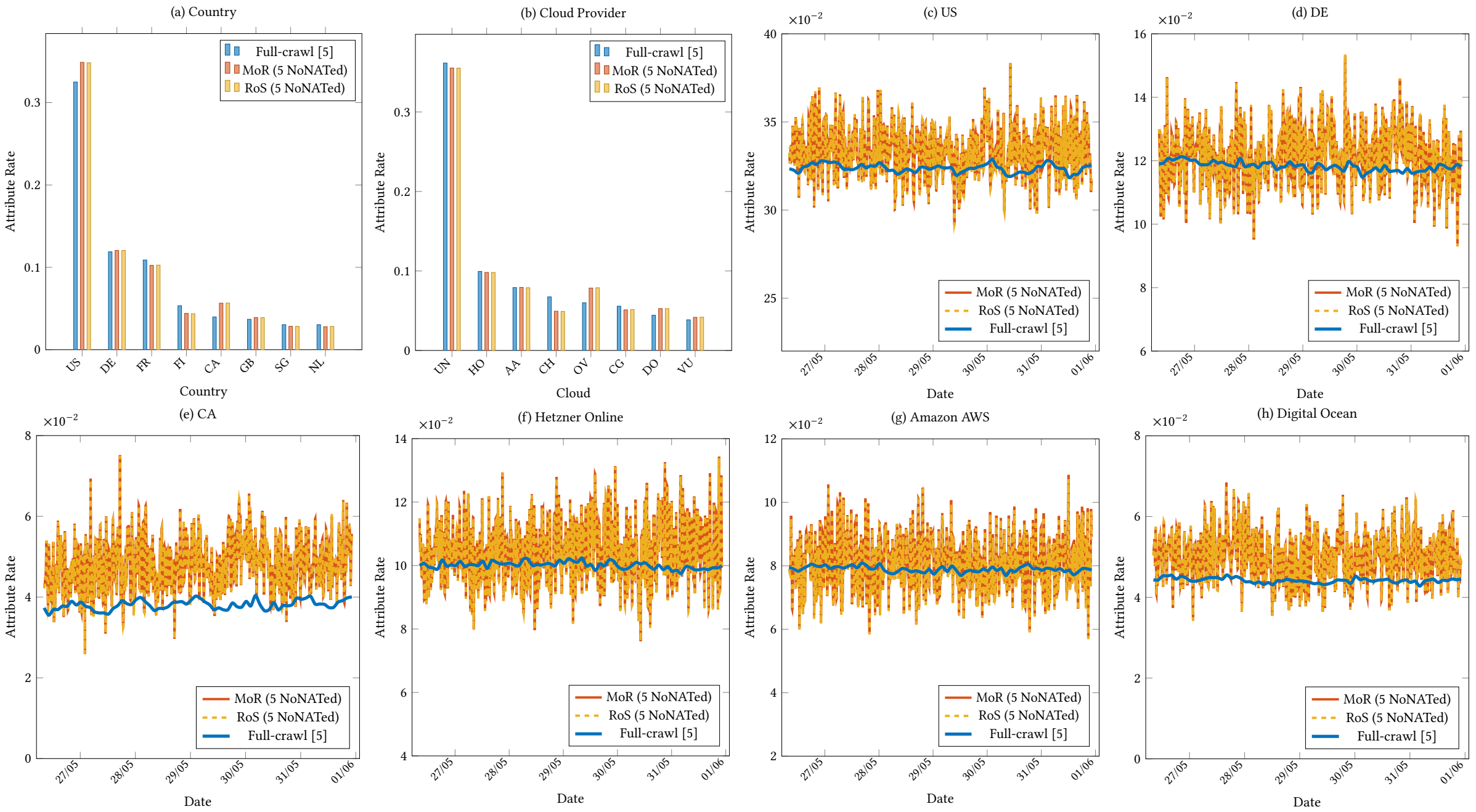

\section{Nodes with and without NAT}\label{sec:app}
%\FloatBarrier

Figures~\ref{fig:nat}(a) and~\ref{fig:nat}(b) compare the attribute prevalences obtained from a full-network crawl with the corresponding MoR and RoS estimates derived from only five NATed nodes within a single snapshot of the Bucket Dataset. These nodes operate as DHT clients and are therefore not directly observable through conventional DHT crawling. Although moderate deviations are observed, the estimated prevalences preserve the relative proportions and ranking of the attributes. This result suggests that the proposed measurement and estimation framework based on NSUM remains effective regardless of whether the measurement nodes are publicly reachable or operate behind NAT.

Figures~\ref{fig:nat}(c)-\ref{fig:nat}(h) evaluate the temporal evolution of different network features considering only five NATed IPFS nodes. Although relying on a very limited number of NATed nodes, both estimators capture the overall temporal behavior of the monitored attributes. The NSUM estimates exhibit substantially larger fluctuations than time series derived from full-network crawls. Since the estimators operate on a much smaller set of observations, they are therefore more sensitive to sampling variability. A notable exception is Amazon AWS, for which both MoR and RoS consistently overestimate the prevalence relative to the curve obtained from crawls, suggesting the presence of attribute-specific visibility biases. These results show that even a very small set of NATed nodes provides sufficient information to monitor the temporal changes in attribute prevalences. Although estimation variance increases compared with the experiments using 10 nodes, the framework remains capable of tracking network dynamics without requiring publicly reachable measurement infrastructure or full-network crawls.

Figures~\ref{fig:nonat}(a) and~\ref{fig:nonat}(b) show the attribute prevalences from a full-network crawl, along with the corresponding MoR and RoS estimates derived from only five nodes without NAT, in a single snapshot of the Bucket Dataset. As shown, the estimated prevalences preserve the relative rates and ranking of the attributes for this snapshot. Figures~\ref{fig:nonat}(c)--\ref{fig:nonat}(h) depict the temporal behavior of the proposed NSUM-based framework when measurements are collected from only five publicly reachable nodes. Although these results are obtained from only five nodes, both estimators capture the overall temporal evolution of the monitored attributes. For all evaluated countries and cloud providers, the estimated prevalences remain close to the curve obtained from full-network crawls and follow the same temporal patterns. Furthermore, the NSUM estimates exhibit larger fluctuation than both the curves obtained from crawls and those obtained from 10 nodes. However, the fluctuations are mainly around the prevalences derived from full-network crawls, indicating that the primary source of error is increased estimation variability due to a small observation set, rather than significant estimation bias.

Unlike the NATed-node experiment shown in Fig.~\ref{fig:nat}(g), no overestimation is observed for Amazon AWS. Instead, the MoR and RoS estimates remain approximately centered around the curve obtained from crawls throughout the measurement period. This suggests that publicly reachable nodes may provide a more representative view of the DHT, although the overall behavior remains consistent across both measurement scenarios. The results indicate that the proposed framework remains effective even when operating with only five publicly reachable nodes. Although the estimation variance is higher than in experiments using larger sample sizes, the framework accurately tracks the temporal dynamics of network attributes without requiring full-network crawls. Furthermore, these results support the feasibility of the proposed measurement and monitoring approach for large-scale Kademlia-based networks.

\end{document}

%% file: acronyms.tex
\begin{acronym}
    \acro{nsum}[NSUM]{the Network Scale-up Method}
    \acro{p2p}[P2P]{peer-to-peer}
    \acro{dht}[DHT]{distributed hash table}
    \acro{ipfs}[IPFS]{the InterPlanetary File System}
    \acro{cid}[CID]{content identifier}
    \acro{ens}[ENS]{the Ethereum Name System}
    \acro{nft}[NFT]{non-fungible token}
    \acro{nb}[NB]{normalized bias}
    \acro{rpc}[RPC]{remote procedure call}
    \acro{nat}[NAT]{network address translation}
\end{acronym}

%% file: figtex/nb_plots.tex
\definecolor{mycolor1}{rgb}{0.00000,0.44700,0.74100}
\definecolor{mycolor2}{rgb}{0.85000,0.32500,0.09800}
\definecolor{mycolor3}{rgb}{0.92900,0.69400,0.12500}
\definecolor{mycolor4}{rgb}{0.46667,0.67451,0.18824}
\definecolor{mycolor5}{rgb}{0.49412,0.18431,0.30588}
\definecolor{mycolor6}{rgb}{0.30196,0.69020,0.93333}

\resizebox{\linewidth}{!}{
\begin{tabular}{c c c c c c}
\hspace{-12pt}
\begin{minipage}{0.20\linewidth}
\resizebox{\linewidth}{!}{
 \begin{tikzpicture}
 	\begin{axis}[
 	legend cell align = left,
    scale = 3.50,
    width = \textwidth,
    height = .9\textwidth,
 	legend pos = north east,
  	grid = both,
    major grid style={gray!25},
    minor grid style={densely dotted, gray!25},
 	ylabel = $\textrm{NB}$,
	xlabel = $\textrm{Sample Size}$,
    % ymajorticks = false,
 	ymin = 0,
 	ymax = 0.05,
 	xmin = 5,
	xmax = 1000,
    xmode = log,
    % title = \textrm{(b) Outage Probability with $r=0.1$m, $\lambda_{obs}=0.5$m$^{-2}$, $d_0 = 10$m},
    title = \textrm{(a) US (MoR)},
 	ticklabel style = {font=\small},
  	]

    \addplot[smooth, line width = 1.50pt, mycolor1] table[x=sample_size, y=median, col sep=comma]{figdata/nb_mor_us.csv};
    \addlegendentry{\footnotesize median}

    \addlegendimage{area legend, fill=mycolor1!20}
    \addlegendentry{\footnotesize p10-p90}

    \addlegendimage{area legend, fill=mycolor1!40}
    \addlegendentry{\footnotesize p25-p75}

    \addplot[name path=lower2, fill=none, smooth, draw=none] table[x=sample_size, y=Q10, col sep=comma]{figdata/nb_mor_us.csv};
    \addplot[name path=upper2, fill=none, smooth, draw=none] table[x=sample_size, y=Q90, col sep=comma]{figdata/nb_mor_us.csv};
    \addplot [mycolor1!20] fill between [of=lower2 and upper2];

    \addplot[name path=lower2, fill=none, smooth, draw=none] table[x=sample_size, y=Q25, col sep=comma]{figdata/nb_mor_us.csv};
    \addplot[name path=upper2, fill=none, smooth, draw=none] table[x=sample_size, y=Q75, col sep=comma]{figdata/nb_mor_us.csv};
    \addplot [mycolor1!40] fill between [of=lower2 and upper2];
    
 	\end{axis}
 \end{tikzpicture}
}
\end{minipage}
&
\hspace{-15pt}
\begin{minipage}{0.20\linewidth}
\resizebox{\linewidth}{!}{
 \begin{tikzpicture}
 	\begin{axis}[
 	legend cell align = left,
    scale = 3.50,
    width = \textwidth,
    height = .9\textwidth,
 	legend pos = south east,
  	grid = both,
    major grid style={gray!25},
    minor grid style={densely dotted, gray!25},
 	ylabel = $\textrm{NB}$,
	xlabel = $\textrm{Sample Size}$,
    % ymajorticks = false,
 	ymin = 0,
 	ymax = 0.12,
 	xmin = 5,
	xmax = 1000,
    xmode = log,
    % title = \textrm{(b) Outage Probability with $r=0.1$m, $\lambda_{obs}=0.5$m$^{-2}$, $d_0 = 10$m},
    title = \textrm{(b) DE (MoR)},
    scaled y ticks=base 10:2,
    ytick scale label code/.code={$\,\times 10^{-2}$},
 	ticklabel style = {font=\small},
  	]

    \addplot[smooth, line width = 1.50pt, mycolor1] table[x=sample_size, y=median, col sep=comma]{figdata/nb_mor_de.csv};

    \addplot[name path=lower2, fill=none, smooth, draw=none] table[x=sample_size, y=Q10, col sep=comma]{figdata/nb_mor_de.csv};
    \addplot[name path=upper2, fill=none, smooth, draw=none] table[x=sample_size, y=Q90, col sep=comma]{figdata/nb_mor_de.csv};
    \addplot [mycolor1!20] fill between [of=lower2 and upper2];
    
    \addplot[name path=lower2, fill=none, smooth, draw=none] table[x=sample_size, y=Q25, col sep=comma]{figdata/nb_mor_de.csv};
    \addplot[name path=upper2, fill=none, smooth, draw=none] table[x=sample_size, y=Q75, col sep=comma]{figdata/nb_mor_de.csv};
    \addplot [mycolor1!40] fill between [of=lower2 and upper2];
    
 	\end{axis}
 \end{tikzpicture}
}
\end{minipage}
&
\hspace{-15pt}
\begin{minipage}{0.20\linewidth}
\resizebox{\linewidth}{!}{
 \begin{tikzpicture}
 	\begin{axis}[
 	legend cell align = left,
    scale = 3.50,
    width = \textwidth,
    height = .9\textwidth,
 	legend pos = south east,
  	grid = both,
    major grid style={gray!25},
    minor grid style={densely dotted, gray!25},
 	ylabel = $\textrm{NB}$,
	xlabel = $\textrm{Sample Size}$,
    % ymajorticks = false,
 	ymin = 0,
 	ymax = 0.2,
 	xmin = 5,
	xmax = 1000,
    xmode = log,
    % title = \textrm{(b) Outage Probability with $r=0.1$m, $\lambda_{obs}=0.5$m$^{-2}$, $d_0 = 10$m},
    title = \textrm{(c) CA (MoR)},
    scaled y ticks=base 10:2,
    ytick scale label code/.code={$\,\times 10^{-2}$},
 	ticklabel style = {font=\small},
  	]

    \addplot[smooth, line width = 1.50pt, mycolor1] table[x=sample_size, y=median, col sep=comma]{figdata/nb_mor_ca.csv};

    \addplot[name path=lower2, fill=none, smooth, draw=none] table[x=sample_size, y=Q10, col sep=comma]{figdata/nb_mor_ca.csv};
    \addplot[name path=upper2, fill=none, smooth, draw=none] table[x=sample_size, y=Q90, col sep=comma]{figdata/nb_mor_ca.csv};
    \addplot [mycolor1!20] fill between [of=lower2 and upper2];
    
    \addplot[name path=lower2, fill=none, smooth, draw=none] table[x=sample_size, y=Q25, col sep=comma]{figdata/nb_mor_ca.csv};
    \addplot[name path=upper2, fill=none, smooth, draw=none] table[x=sample_size, y=Q75, col sep=comma]{figdata/nb_mor_ca.csv};
    \addplot [mycolor1!40] fill between [of=lower2 and upper2];
    
 	\end{axis}
 \end{tikzpicture}
}
\end{minipage}
&
\hspace{-15pt}
\begin{minipage}{0.20\linewidth}
\resizebox{\linewidth}{!}{
 \begin{tikzpicture}
 	\begin{axis}[
 	legend cell align = left,
    scale = 3.50,
    width = \textwidth,
    height = .9\textwidth,
 	legend pos = north east,
  	grid = both,
    major grid style={gray!25},
    minor grid style={densely dotted, gray!25},
 	ylabel = $\textrm{NB}$,
	xlabel = $\textrm{Sample Size}$,
    % ymajorticks = false,
 	ymin = 0,
 	ymax = 0.12,
 	xmin = 5,
	xmax = 1000,
    xmode = log,
    % title = \textrm{(b) Outage Probability with $r=0.1$m, $\lambda_{obs}=0.5$m$^{-2}$, $d_0 = 10$m},
    title = \textrm{(d) Hetzner Online (MoR)},
    scaled y ticks=base 10:2,
    ytick scale label code/.code={$\,\times 10^{-2}$},
 	ticklabel style = {font=\small},
  	]

    \addplot[smooth, line width = 1.50pt, mycolor1] table[x=sample_size, y=median, col sep=comma]{figdata/nb_mor_ho.csv};
    % \addlegendentry{\scriptsize \texttt{median}}

    % \addlegendimage{area legend, fill=mycolor2!20}
    % \addlegendentry{\scriptsize \texttt{p10-p90}}

    % \addlegendimage{area legend, fill=mycolor2!40}
    % \addlegendentry{\scriptsize \texttt{p25-p75}}

    \addplot[name path=lower2, fill=none, smooth, draw=none] table[x=sample_size, y=Q10, col sep=comma]{figdata/nb_mor_ho.csv};
    \addplot[name path=upper2, fill=none, smooth, draw=none] table[x=sample_size, y=Q90, col sep=comma]{figdata/nb_mor_ho.csv};
    \addplot [mycolor1!20] fill between [of=lower2 and upper2];
    
    \addplot[name path=lower2, fill=none, smooth, draw=none] table[x=sample_size, y=Q25, col sep=comma]{figdata/nb_mor_ho.csv};
    \addplot[name path=upper2, fill=none, smooth, draw=none] table[x=sample_size, y=Q75, col sep=comma]{figdata/nb_mor_ho.csv};
    \addplot [mycolor1!40] fill between [of=lower2 and upper2];
    
 	\end{axis}
 \end{tikzpicture}
}
\end{minipage}
&
\hspace{-15pt}
\begin{minipage}{0.20\linewidth}
\resizebox{\linewidth}{!}{
 \begin{tikzpicture}
 	\begin{axis}[
 	legend cell align = left,
    scale = 3.50,
    width = \textwidth,
    height = .9\textwidth,
 	legend pos = south east,
  	grid = both,
    major grid style={gray!25},
    minor grid style={densely dotted, gray!25},
 	ylabel = $\textrm{NB}$,
	xlabel = $\textrm{Sample Size}$,
    % ymajorticks = false,
 	ymin = 0,
 	ymax = 0.15,
 	xmin = 5,
	xmax = 1000,
    xmode = log,
    % title = \textrm{(b) Outage Probability with $r=0.1$m, $\lambda_{obs}=0.5$m$^{-2}$, $d_0 = 10$m},
    title = \textrm{(e) Amazon AWS (MoR)},
    scaled y ticks=base 10:2,
    ytick scale label code/.code={$\,\times 10^{-2}$},
 	ticklabel style = {font=\small},
  	]

    \addplot[smooth, line width = 1.50pt, mycolor1] table[x=sample_size, y=median, col sep=comma]{figdata/nb_mor_aa.csv};

    \addplot[name path=lower2, fill=none, smooth, draw=none] table[x=sample_size, y=Q10, col sep=comma]{figdata/nb_mor_aa.csv};
    \addplot[name path=upper2, fill=none, smooth, draw=none] table[x=sample_size, y=Q90, col sep=comma]{figdata/nb_mor_aa.csv};
    \addplot [mycolor1!20] fill between [of=lower2 and upper2];
    
    \addplot[name path=lower2, fill=none, smooth, draw=none] table[x=sample_size, y=Q25, col sep=comma]{figdata/nb_mor_aa.csv};
    \addplot[name path=upper2, fill=none, smooth, draw=none] table[x=sample_size, y=Q75, col sep=comma]{figdata/nb_mor_aa.csv};
    \addplot [mycolor1!40] fill between [of=lower2 and upper2];
    
 	\end{axis}
 \end{tikzpicture}
}
\end{minipage}
&
\hspace{-15pt}
\begin{minipage}{0.20\linewidth}
\resizebox{\linewidth}{!}{
 \begin{tikzpicture}
 	\begin{axis}[
 	legend cell align = left,
    scale = 3.50,
    width = \textwidth,
    height = .9\textwidth,
 	legend pos = south east,
  	grid = both,
    major grid style={gray!25},
    minor grid style={densely dotted, gray!25},
 	ylabel = $\textrm{NB}$,
	xlabel = $\textrm{Sample Size}$,
    % ymajorticks = false,
 	ymin = 0,
 	ymax = 0.15,
 	xmin = 5,
	xmax = 1000,
    xmode = log,
    % title = \textrm{(b) Outage Probability with $r=0.1$m, $\lambda_{obs}=0.5$m$^{-2}$, $d_0 = 10$m},
    title = \textrm{(f) Digital Ocean (MoR)},
    scaled y ticks=base 10:2,
    ytick scale label code/.code={$\,\times 10^{-2}$},
 	ticklabel style = {font=\small},
  	]

    \addplot[smooth, line width = 1.50pt, mycolor1] table[x=sample_size, y=median, col sep=comma]{figdata/nb_mor_do.csv};

    \addplot[name path=lower2, fill=none, smooth, draw=none] table[x=sample_size, y=Q10, col sep=comma]{figdata/nb_mor_do.csv};
    \addplot[name path=upper2, fill=none, smooth, draw=none] table[x=sample_size, y=Q90, col sep=comma]{figdata/nb_mor_do.csv};
    \addplot [mycolor1!20] fill between [of=lower2 and upper2];
    
    \addplot[name path=lower2, fill=none, smooth, draw=none] table[x=sample_size, y=Q25, col sep=comma]{figdata/nb_mor_do.csv};
    \addplot[name path=upper2, fill=none, smooth, draw=none] table[x=sample_size, y=Q75, col sep=comma]{figdata/nb_mor_do.csv};
    \addplot [mycolor1!40] fill between [of=lower2 and upper2];
    
 	\end{axis}
 \end{tikzpicture}
}
\end{minipage}
\\
\hspace{-12pt}
\begin{minipage}{0.20\linewidth}
\resizebox{\linewidth}{!}{
 \begin{tikzpicture}
 	\begin{axis}[
 	legend cell align = left,
    scale = 3.50,
    width = \textwidth,
    height = .9\textwidth,
 	legend pos = south east,
  	grid = both,
    major grid style={gray!25},
    minor grid style={densely dotted, gray!25},
 	ylabel = $\textrm{NB}$,
	xlabel = $\textrm{Sample Size}$,
    % ymajorticks = false,
 	ymin = 0,
 	ymax = 0.05,
 	xmin = 5,
	xmax = 1000,
    xmode = log,
    % title = \textrm{(b) Outage Probability with $r=0.1$m, $\lambda_{obs}=0.5$m$^{-2}$, $d_0 = 10$m},
    title = \textrm{(g) US (RoS)},
    scaled y ticks=base 10:2,
    ytick scale label code/.code={$\,\times 10^{-2}$},
 	ticklabel style = {font=\small},
  	]

    \addplot[smooth, line width = 1.50pt, mycolor1] table[x=sample_size, y=median, col sep=comma]{figdata/nb_ros_us.csv};

    \addplot[name path=lower2, fill=none, smooth, draw=none] table[x=sample_size, y=Q10, col sep=comma]{figdata/nb_ros_us.csv};
    \addplot[name path=upper2, fill=none, smooth, draw=none] table[x=sample_size, y=Q90, col sep=comma]{figdata/nb_ros_us.csv};
    \addplot [mycolor1!20] fill between [of=lower2 and upper2];
    
    \addplot[name path=lower2, fill=none, smooth, draw=none] table[x=sample_size, y=Q25, col sep=comma]{figdata/nb_ros_us.csv};
    \addplot[name path=upper2, fill=none, smooth, draw=none] table[x=sample_size, y=Q75, col sep=comma]{figdata/nb_ros_us.csv};
    \addplot [mycolor1!40] fill between [of=lower2 and upper2];
    
 	\end{axis}
 \end{tikzpicture}
}
\end{minipage}
&
\hspace{-15pt}
\begin{minipage}{0.20\linewidth}
\resizebox{\linewidth}{!}{
 \begin{tikzpicture}
 	\begin{axis}[
 	legend cell align = left,
    scale = 3.50,
    width = \textwidth,
    height = .9\textwidth,
 	legend pos = south east,
  	grid = both,
    major grid style={gray!25},
    minor grid style={densely dotted, gray!25},
 	ylabel = $\textrm{NB}$,
	xlabel = $\textrm{Sample Size}$,
    % ymajorticks = false,
 	ymin = 0,
 	ymax = 0.12,
 	xmin = 5,
	xmax = 1000,
    xmode = log,
    % title = \textrm{(b) Outage Probability with $r=0.1$m, $\lambda_{obs}=0.5$m$^{-2}$, $d_0 = 10$m},
    title = \textrm{(h) DE (RoS)},
    scaled y ticks=base 10:2,
    ytick scale label code/.code={$\,\times 10^{-2}$},
 	ticklabel style = {font=\small},
  	]

    \addplot[smooth, line width = 1.50pt, mycolor1] table[x=sample_size, y=median, col sep=comma]{figdata/nb_ros_de.csv};

    \addplot[name path=lower2, fill=none, smooth, draw=none] table[x=sample_size, y=Q10, col sep=comma]{figdata/nb_ros_de.csv};
    \addplot[name path=upper2, fill=none, smooth, draw=none] table[x=sample_size, y=Q90, col sep=comma]{figdata/nb_ros_de.csv};
    \addplot [mycolor1!20] fill between [of=lower2 and upper2];
    
    \addplot[name path=lower2, fill=none, smooth, draw=none] table[x=sample_size, y=Q25, col sep=comma]{figdata/nb_ros_de.csv};
    \addplot[name path=upper2, fill=none, smooth, draw=none] table[x=sample_size, y=Q75, col sep=comma]{figdata/nb_ros_de.csv};
    \addplot [mycolor1!40] fill between [of=lower2 and upper2];
    
 	\end{axis}
 \end{tikzpicture}
}
\end{minipage}
&
\hspace{-15pt}
\begin{minipage}{0.20\linewidth}
\resizebox{\linewidth}{!}{
 \begin{tikzpicture}
 	\begin{axis}[
 	legend cell align = left,
    scale = 3.50,
    width = \textwidth,
    height = .9\textwidth,
 	legend pos = south east,
  	grid = both,
    major grid style={gray!25},
    minor grid style={densely dotted, gray!25},
 	ylabel = $\textrm{NB}$,
	xlabel = $\textrm{Sample Size}$,
    % ymajorticks = false,
 	ymin = 0,
 	ymax = 0.2,
 	xmin = 5,
	xmax = 1000,
    xmode = log,
    % title = \textrm{(b) Outage Probability with $r=0.1$m, $\lambda_{obs}=0.5$m$^{-2}$, $d_0 = 10$m},
    title = \textrm{(i) CA (RoS)},
    scaled y ticks=base 10:2,
    ytick scale label code/.code={$\,\times 10^{-2}$},
 	ticklabel style = {font=\small},
  	]

    \addplot[smooth, line width = 1.50pt, mycolor1] table[x=sample_size, y=median, col sep=comma]{figdata/nb_ros_ca.csv};

    \addplot[name path=lower2, fill=none, smooth, draw=none] table[x=sample_size, y=Q10, col sep=comma]{figdata/nb_ros_ca.csv};
    \addplot[name path=upper2, fill=none, smooth, draw=none] table[x=sample_size, y=Q90, col sep=comma]{figdata/nb_ros_ca.csv};
    \addplot [mycolor1!20] fill between [of=lower2 and upper2];
    
    \addplot[name path=lower2, fill=none, smooth, draw=none] table[x=sample_size, y=Q25, col sep=comma]{figdata/nb_ros_ca.csv};
    \addplot[name path=upper2, fill=none, smooth, draw=none] table[x=sample_size, y=Q75, col sep=comma]{figdata/nb_ros_ca.csv};
    \addplot [mycolor1!40] fill between [of=lower2 and upper2];
    
 	\end{axis}
 \end{tikzpicture}
}
\end{minipage}
&
\hspace{-15pt}
\begin{minipage}{0.20\linewidth}
\resizebox{\linewidth}{!}{
 \begin{tikzpicture}
 	\begin{axis}[
 	legend cell align = left,
    scale = 3.50,
    width = \textwidth,
    height = .9\textwidth,
 	legend pos = south east,
  	grid = both,
    major grid style={gray!25},
    minor grid style={densely dotted, gray!25},
 	ylabel = $\textrm{NB}$,
	xlabel = $\textrm{Sample Size}$,
    % ymajorticks = false,
 	ymin = 0,
 	ymax = 0.12,
 	xmin = 5,
	xmax = 1000,
    xmode = log,
    % title = \textrm{(b) Outage Probability with $r=0.1$m, $\lambda_{obs}=0.5$m$^{-2}$, $d_0 = 10$m},
    title = \textrm{(j) Heztner Online (RoS)},
    scaled y ticks=base 10:2,
    ytick scale label code/.code={$\,\times 10^{-2}$},
 	ticklabel style = {font=\small},
  	]

    \addplot[smooth, line width = 1.50pt, mycolor1] table[x=sample_size, y=median, col sep=comma]{figdata/nb_ros_ho.csv};

    \addplot[name path=lower2, fill=none, smooth, draw=none] table[x=sample_size, y=Q10, col sep=comma]{figdata/nb_ros_ho.csv};
    \addplot[name path=upper2, fill=none, smooth, draw=none] table[x=sample_size, y=Q90, col sep=comma]{figdata/nb_ros_ho.csv};
    \addplot [mycolor1!20] fill between [of=lower2 and upper2];
    
    \addplot[name path=lower2, fill=none, smooth, draw=none] table[x=sample_size, y=Q25, col sep=comma]{figdata/nb_ros_ho.csv};
    \addplot[name path=upper2, fill=none, smooth, draw=none] table[x=sample_size, y=Q75, col sep=comma]{figdata/nb_ros_ho.csv};
    \addplot [mycolor1!40] fill between [of=lower2 and upper2];
    
 	\end{axis}
 \end{tikzpicture}
}
\end{minipage}
&
\hspace{-15pt}
\begin{minipage}{0.20\linewidth}
\resizebox{\linewidth}{!}{
 \begin{tikzpicture}
 	\begin{axis}[
 	legend cell align = left,
    scale = 3.50,
    width = \textwidth,
    height = .9\textwidth,
 	legend pos = south east,
  	grid = both,
    major grid style={gray!25},
    minor grid style={densely dotted, gray!25},
 	ylabel = $\textrm{NB}$,
	xlabel = $\textrm{Sample Size}$,
    % ymajorticks = false,
 	ymin = 0,
 	ymax = 0.15,
 	xmin = 5,
	xmax = 1000,
    xmode = log,
    % title = \textrm{(b) Outage Probability with $r=0.1$m, $\lambda_{obs}=0.5$m$^{-2}$, $d_0 = 10$m},
    title = \textrm{(k) Amazon AWS (RoS)},
    scaled y ticks=base 10:2,
    ytick scale label code/.code={$\,\times 10^{-2}$},
 	ticklabel style = {font=\small},
  	]

    \addplot[smooth, line width = 1.50pt, mycolor1] table[x=sample_size, y=median, col sep=comma]{figdata/nb_ros_aa.csv};

    \addplot[name path=lower2, fill=none, smooth, draw=none] table[x=sample_size, y=Q10, col sep=comma]{figdata/nb_ros_aa.csv};
    \addplot[name path=upper2, fill=none, smooth, draw=none] table[x=sample_size, y=Q90, col sep=comma]{figdata/nb_ros_aa.csv};
    \addplot [mycolor1!20] fill between [of=lower2 and upper2];
    
    \addplot[name path=lower2, fill=none, smooth, draw=none] table[x=sample_size, y=Q25, col sep=comma]{figdata/nb_ros_aa.csv};
    \addplot[name path=upper2, fill=none, smooth, draw=none] table[x=sample_size, y=Q75, col sep=comma]{figdata/nb_ros_aa.csv};
    \addplot [mycolor1!40] fill between [of=lower2 and upper2];
    
 	\end{axis}
 \end{tikzpicture}
}
\end{minipage}
&
\hspace{-15pt}
\begin{minipage}{0.20\linewidth}
\resizebox{\linewidth}{!}{
 \begin{tikzpicture}
 	\begin{axis}[
 	legend cell align = left,
    scale = 3.50,
    width = \textwidth,
    height = .9\textwidth,
 	legend pos = south east,
  	grid = both,
    major grid style={gray!25},
    minor grid style={densely dotted, gray!25},
 	ylabel = $\textrm{NB}$,
	xlabel = $\textrm{Sample Size}$,
    % ymajorticks = false,
 	ymin = 0,
 	ymax = 0.15,
 	xmin = 5,
	xmax = 1000,
    xmode = log,
    % title = \textrm{(b) Outage Probability with $r=0.1$m, $\lambda_{obs}=0.5$m$^{-2}$, $d_0 = 10$m},
    title = \textrm{(l) Digital Ocean (RoS)},
    scaled y ticks=base 10:2,
    ytick scale label code/.code={$\,\times 10^{-2}$},
 	ticklabel style = {font=\small},
  	]

    \addplot[smooth, line width = 1.50pt, mycolor1] table[x=sample_size, y=median, col sep=comma]{figdata/nb_ros_do.csv};

    \addplot[name path=lower2, fill=none, smooth, draw=none] table[x=sample_size, y=Q10, col sep=comma]{figdata/nb_ros_do.csv};
    \addplot[name path=upper2, fill=none, smooth, draw=none] table[x=sample_size, y=Q90, col sep=comma]{figdata/nb_ros_do.csv};
    \addplot [mycolor1!20] fill between [of=lower2 and upper2];
    
    \addplot[name path=lower2, fill=none, smooth, draw=none] table[x=sample_size, y=Q25, col sep=comma]{figdata/nb_ros_do.csv};
    \addplot[name path=upper2, fill=none, smooth, draw=none] table[x=sample_size, y=Q75, col sep=comma]{figdata/nb_ros_do.csv};
    \addplot [mycolor1!40] fill between [of=lower2 and upper2];
    
 	\end{axis}
 \end{tikzpicture}
}
\end{minipage}
\end{tabular} 
}
\vspace{-10pt}

%% file: figtex/bckt_results.tex
\definecolor{mycolor1}{rgb}{0.00000,0.44700,0.74100}
\definecolor{mycolor2}{rgb}{0.85000,0.32500,0.09800}
\definecolor{mycolor3}{rgb}{0.92900,0.69400,0.12500}
\definecolor{mycolor4}{rgb}{0.46667,0.67451,0.18824}
\definecolor{mycolor5}{rgb}{0.49412,0.18431,0.30588}
\definecolor{mycolor6}{rgb}{0.30196,0.69020,0.93333}

\resizebox{.9\linewidth}{!}{
\begin{tabular}{c c c c c}
\hspace{-12pt}
\begin{minipage}{0.25\linewidth}
\resizebox{\linewidth}{!}{
\pgfplotstableread[col sep=comma]{figdata/country_stats_round_585.csv}\datatable
\begin{tikzpicture}
\begin{axis}[
    ybar,
    bar width=2.30pt,
    scale = 2.50,
    width=\textwidth,
    height=\textwidth,
    xlabel={Country},
    ylabel={Attribute Rate},
    title = {(a) Country},
    % symbolic x coords={}, % not needed; row index is used internally
    xtick=data,
    xticklabels from table={\datatable}{country},
    x tick label style={
        rotate=45,
        anchor=east,
        font=\small
    },
    enlarge x limits=0.15,
    ymin=0,
    % legend style={
    %     at={(0.5,1.05)},
    %     anchor=south,
    %     legend columns=3
    % },
    % nodes near coords,
    % nodes near coords align={vertical},
]

\addplot+[
    fill=mycolor1!60,
    draw=mycolor1!80!black
    ] table[
    x expr=\coordindex,
    y=props_gt
]{\datatable};

\addplot+[
    fill=mycolor2!60,
    draw=mycolor2!80!black,
    error bars/.cd,
    y dir=both,
    y explicit,
    error bar style={black, thick}
    ] table[
    x expr=\coordindex,
    y=mor_mean,
    y error = mor_ci_95
]{\datatable};

\addplot+[
    fill=mycolor3!60,
    draw=mycolor3!80!black,
    error bars/.cd,
    y dir=both,
    y explicit,
    error bar style={black, thick}
    ] table[
    x expr=\coordindex,
    y=ros_mean,
    y error = ros_ci_95
]{\datatable};

\legend{Full-crawl \cite{Balduf2023Cloud}, MoR, RoS}

\end{axis}
\end{tikzpicture}
}
\end{minipage}
&
\hspace{-12pt}
\begin{minipage}{0.25\linewidth}
\resizebox{\linewidth}{!}{
\begin{tikzpicture}
\begin{axis}[
    ybar,
    bar width=2.30pt,
    scale = 2.50,
    width=\textwidth,
    height=\textwidth,
    xlabel={Cloud},
    ylabel={Attribute Rate},
    title = {(b) Cloud Provider},
    % symbolic x coords={}, % not needed; row index is used internally
    xtick=data,
    xticklabels={NC, HO, CH, CG, AA, OV, VU, DO},
    x tick label style={
        rotate=45,
        anchor=east,
        font=\small
    },
    enlarge x limits=0.15,
    ymin=0,
    % legend style={
    %     at={(0.5,1.05)},
    %     anchor=south,
    %     legend columns=3
    % },
    % nodes near coords,
    % nodes near coords align={vertical},
]

\addplot+[
    fill=mycolor1!60,
    draw=mycolor1!80!black
    ] table[
    x expr=\coordindex,
    y=props_gt,
    col sep = comma
]{figdata/country_stats_round_585.csv};

\addplot+[
    fill=mycolor2!60,
    draw=mycolor2!80!black,
    error bars/.cd,
    y dir=both,
    y explicit,
    error bar style={black, thick}
    ] table[
    x expr=\coordindex,
    y=mor_mean,
    y error = mor_ci_95,    
    col sep = comma
]{figdata/country_stats_round_585.csv};

\addplot+[
    fill=mycolor3!60,
    draw=mycolor3!80!black,
    error bars/.cd,
    y dir=both,
    y explicit,
    error bar style={black, thick}
    ] table[
    x expr=\coordindex,
    y=ros_mean,
    y error = ros_ci_95,
    col sep = comma
]{figdata/country_stats_round_585.csv};

\legend{Full-crawl \cite{Balduf2023Cloud}, MoR, RoS}

\end{axis}
\end{tikzpicture}
}
\end{minipage}
&
\hspace{-12pt}
\begin{minipage}{0.25\linewidth}
\resizebox{\linewidth}{!}{
\begin{tikzpicture}
\begin{axis}[
    scale = 2.50,
    width=\textwidth,
    height=\textwidth,
    legend pos = south east,
    xlabel={Date},
    ylabel={Attribute Rate},
    title = {(c) US},
    date coordinates in=x,
    xticklabel={\day/\month},
    % symbolic x coords={}, % not needed; row index is used internally
    % xtick=data,
    % xticklabels=\year-\month-\day,
    x tick label style={
        rotate=45,
        anchor=east,
        font=\small
    },
    enlarge x limits=0.025,
    ymin=0.26,
    ymax=0.36,
    scaled y ticks=base 10:2,
    ytick scale label code/.code={$\,\times 10^{-2}$},
    % legend style={
    %     at={(0.5,1.05)},
    %     anchor=south,
    %     legend columns=3
    % },
    % nodes near coords,
    % nodes near coords align={vertical},
]

\addplot[
thick,
mycolor2,
line width=1.50,
] table[
x = datetime,
y = us_mor_sim,
col sep=comma,
]{figdata/country_temporal.csv};
\addlegendentry{MoR}

\addplot[
dashed,
mycolor3,
line width=1.50,
] table[
x = datetime,
y = us_ros_sim,
col sep=comma,
]{figdata/country_temporal.csv};
\addlegendentry{RoS}

\addplot[
thick,
mycolor1,
line width=2,
] table[
x = datetime,
y = us_real_prop,
col sep=comma,
]{figdata/country_temporal.csv};
\addlegendentry{Full-crawl \cite{Balduf2023Cloud}}

\end{axis}
\end{tikzpicture}
}
\end{minipage}
&
\hspace{-12pt}
\begin{minipage}{0.25\linewidth}
\resizebox{\linewidth}{!}{
\begin{tikzpicture}
\begin{axis}[
    scale = 2.50,
    width=\textwidth,
    height=\textwidth,
    legend pos = south east,
    xlabel={Date},
    ylabel={Attribute Rate},
    title = {(d) DE},
    date coordinates in=x,
    xticklabel={\day/\month},
    % symbolic x coords={}, % not needed; row index is used internally
    % xtick=data,
    % xticklabels=\year-\month-\day,
    x tick label style={
        rotate=45,
        anchor=east,
        font=\small
    },
    enlarge x limits=0.025,
    ymin=0.075,
    ymax=0.112,
    scaled y ticks=base 10:2,
    ytick scale label code/.code={$\,\times 10^{-2}$},
    % legend style={
    %     at={(0.5,1.05)},
    %     anchor=south,
    %     legend columns=3
    % },
    % nodes near coords,
    % nodes near coords align={vertical},
]

\addplot[
thick,
mycolor2,
line width=1.50,
] table[
x = datetime,
y = de_mor_sim,
col sep=comma,
]{figdata/country_temporal.csv};
\addlegendentry{MoR}

\addplot[
dashed,
mycolor3,
line width=1.50,
] table[
x = datetime,
y = de_ros_sim,
col sep=comma,
]{figdata/country_temporal.csv};
\addlegendentry{RoS}

\addplot[
thick,
mycolor1,
line width=2,
] table[
x = datetime,
y = de_real_prop,
col sep=comma,
]{figdata/country_temporal.csv};
\addlegendentry{Full-crawl \cite{Balduf2023Cloud}}

\end{axis}
\end{tikzpicture}
}
\end{minipage}
&
\hspace{-12pt}
\begin{minipage}{0.25\linewidth}
\resizebox{\linewidth}{!}{
\begin{tikzpicture}
\begin{axis}[
    scale = 2.50,
    width=\textwidth,
    height=\textwidth,
    legend pos = south east,
    xlabel={Date},
    ylabel={Attribute Rate},
    title = {(e) CA},
    date coordinates in=x,
    xticklabel={\day/\month},
    % symbolic x coords={}, % not needed; row index is used internally
    % xtick=data,
    % xticklabels=\year-\month-\day,
    x tick label style={
        rotate=45,
        anchor=east,
        font=\small
    },
    enlarge x limits=0.025,
    ymin=0.02,
    ymax=0.042,
    scaled y ticks=base 10:2,
    ytick scale label code/.code={$\,\times 10^{-2}$},
    % legend style={
    %     at={(0.5,1.05)},
    %     anchor=south,
    %     legend columns=3
    % },
    % nodes near coords,
    % nodes near coords align={vertical},
]

\addplot[
thick,
mycolor2,
line width=1.50,
] table[
x = datetime,
y = ca_mor_sim,
col sep=comma,
]{figdata/country_temporal.csv};
\addlegendentry{MoR}

\addplot[
dashed,
mycolor3,
line width=1.50,
] table[
x = datetime,
y = ca_ros_sim,
col sep=comma,
]{figdata/country_temporal.csv};
\addlegendentry{RoS}

\addplot[
thick,
mycolor1,
line width=2,
] table[
x = datetime,
y = ca_real_prop,
col sep=comma,
]{figdata/country_temporal.csv};
\addlegendentry{Full-crawl \cite{Balduf2023Cloud}}

\end{axis}
\end{tikzpicture}
}
\end{minipage}
\\
\hspace{-12pt}
\begin{minipage}{0.25\linewidth}
\resizebox{\linewidth}{!}{
\begin{tikzpicture}
\begin{axis}[
    scale = 2.50,
    width=\textwidth,
    height=\textwidth,
    legend pos = south east,
    xlabel={Date},
    ylabel={Attribute Rate},
    title = {(f) Hetzner Online},
    date coordinates in=x,
    xticklabel={\day/\month},
    x tick label style={
        rotate=45,
        anchor=east,
        font=\small
    },
    enlarge x limits=0.025,
    ymin=0.07,
    ymax=0.12,
    scaled y ticks=base 10:2,
    ytick scale label code/.code={$\,\times 10^{-2}$},
]

\addplot[
thick,
mycolor2,
line width=1.50,
] table[
x = datetime,
y = mor_sim,
col sep=comma,
]{figdata/cloud_temporal.csv};
\addlegendentry{MoR}

\addplot[
dashed,
mycolor3,
line width=1.50,
] table[
x = datetime,
y = ros_sim,
col sep=comma,
]{figdata/cloud_temporal.csv};
\addlegendentry{RoS}

\addplot[
thick,
mycolor1,
line width=2,
] table[
x = datetime,
y = real_prop,
col sep=comma,
]{figdata/cloud_temporal.csv};
\addlegendentry{Full-crawl \cite{Balduf2023Cloud}}

\end{axis}
\end{tikzpicture}
}
\end{minipage}
&
\hspace{-12pt}
\begin{minipage}{0.25\linewidth}
\resizebox{\linewidth}{!}{
\begin{tikzpicture}
\begin{axis}[
    scale = 2.50,
    width=\textwidth,
    height=\textwidth,
    legend pos = south east,
    xlabel={Date},
    ylabel={Attribute Rate},
    title = {(g) Amazon AWS},
    date coordinates in=x,
    xticklabel={\day/\month},
    % symbolic x coords={}, % not needed; row index is used internally
    % xtick=data,
    % xticklabels=\year-\month-\day,
    x tick label style={
        rotate=45,
        anchor=east,
        font=\small
    },
    enlarge x limits=0.025,
    ymin=0.05,
    ymax=0.085,
    scaled y ticks=base 10:2,
    ytick scale label code/.code={$\,\times 10^{-2}$},
    % legend style={
    %     at={(0.5,1.05)},
    %     anchor=south,
    %     legend columns=3
    % },
    % nodes near coords,
    % nodes near coords align={vertical},
]

\addplot[
thick,
mycolor2,
line width=1.50,
] table[
x = datetime,
y = am_mor_sim,
col sep=comma,
]{figdata/cloud_temporal.csv};
\addlegendentry{MoR}

\addplot[
dashed,
mycolor3,
line width=1.50,
] table[
x = datetime,
y = am_ros_sim,
col sep=comma,
]{figdata/cloud_temporal.csv};
\addlegendentry{RoS}

\addplot[
thick,
mycolor1,
line width=2,
] table[
x = datetime,
y = am_real_prop,
col sep=comma,
]{figdata/cloud_temporal.csv};
\addlegendentry{Full-crawl \cite{Balduf2023Cloud}}

\end{axis}
\end{tikzpicture}
}
\end{minipage}
&
\hspace{-12pt}
\begin{minipage}{0.25\linewidth}
\resizebox{\linewidth}{!}{
\begin{tikzpicture}
\begin{axis}[
    scale = 2.50,
    width=\textwidth,
    height=\textwidth,
    legend pos = south east,
    xlabel={Date},
    ylabel={Attribute Rate},
    title = {(h) Digital Ocean},
    date coordinates in=x,
    xticklabel={\day/\month},
    % symbolic x coords={}, % not needed; row index is used internally
    % xtick=data,
    % xticklabels=\year-\month-\day,
    x tick label style={
        rotate=45,
        anchor=east,
        font=\small
    },
    enlarge x limits=0.025,
    ymin=0.028,
    ymax=0.055,
    scaled y ticks=base 10:2,
    ytick scale label code/.code={$\,\times 10^{-2}$},
    % legend style={
    %     at={(0.5,1.05)},
    %     anchor=south,
    %     legend columns=3
    % },
    % nodes near coords,
    % nodes near coords align={vertical},
]

\addplot[
thick,
mycolor2,
line width=1.50,
] table[
x = datetime,
y = do_mor_sim,
col sep=comma,
]{figdata/cloud_temporal.csv};
\addlegendentry{MoR}

\addplot[
dashed,
mycolor3,
line width=1.50,
] table[
x = datetime,
y = do_ros_sim,
col sep=comma,
]{figdata/cloud_temporal.csv};
\addlegendentry{RoS}

\addplot[
thick,
mycolor1,
line width=2,
] table[
x = datetime,
y = do_real_prop,
col sep=comma,
]{figdata/cloud_temporal.csv};
\addlegendentry{Full-crawl \cite{Balduf2023Cloud}}

\end{axis}
\end{tikzpicture}
}
\end{minipage}
&
\hspace{-12pt}
\begin{minipage}{0.25\linewidth}
\resizebox{\linewidth}{!}{
\begin{tikzpicture}
\begin{axis}[
    scale = 2.50,
    width=\textwidth,
    height=\textwidth,
    legend pos = south east,
    xlabel={Date},
    ylabel={Attribute Rate},
    title = {(i) Non-cloud},
    date coordinates in=x,
    xticklabel={\day/\month},
    % symbolic x coords={}, % not needed; row index is used internally
    % xtick=data,
    % xticklabels=\year-\month-\day,
    x tick label style={
        rotate=45,
        anchor=east,
        font=\small
    },
    enlarge x limits=0.025,
    ymin=0.30,
    ymax=0.40,
    scaled y ticks=base 10:2,
    ytick scale label code/.code={$\,\times 10^{-2}$},
    % legend style={
    %     at={(0.5,1.05)},
    %     anchor=south,
    %     legend columns=3
    % },
    % nodes near coords,
    % nodes near coords align={vertical},
]

\addplot[
thick,
mycolor2,
line width=1.50,
] table[
x = datetime,
y = unkn_mor_sim,
col sep=comma,
]{figdata/cloud_temporal.csv};
\addlegendentry{MoR}

\addplot[
dashed,
mycolor3,
line width=1.50,
] table[
x = datetime,
y = unkn_ros_sim,
col sep=comma,
]{figdata/cloud_temporal.csv};
\addlegendentry{RoS}

\addplot[
thick,
mycolor1,
line width=2,
] table[
x = datetime,
y = unkn_real_prop,
col sep=comma,
]{figdata/cloud_temporal.csv};
\addlegendentry{Full-crawl \cite{Balduf2023Cloud}}

\end{axis}
\end{tikzpicture}
}
\end{minipage}
&
\hspace{-12pt}
\begin{minipage}{0.25\linewidth}
\resizebox{\linewidth}{!}{
\begin{tikzpicture}
\begin{axis}[
    scale = 2.50,
    width=\textwidth,
    height=\textwidth,
    legend pos = north east,
    xlabel={Date},
    ylabel={Attribute Rate},
    title = {(j) Cloud},
    date coordinates in=x,
    xticklabel={\day/\month},
    % symbolic x coords={}, % not needed; row index is used internally
    % xtick=data,
    % xticklabels=\year-\month-\day,
    x tick label style={
        rotate=45,
        anchor=east,
        font=\small
    },
    enlarge x limits=0.025,
    ymin=0.60,
    ymax=0.70,
    scaled y ticks=base 10:2,
    ytick scale label code/.code={$\,\times 10^{-2}$},
    % legend style={
    %     at={(0.5,1.05)},
    %     anchor=south,
    %     legend columns=3
    % },
    % nodes near coords,
    % nodes near coords align={vertical},
]

\addplot[
thick,
mycolor2,
line width=1.50,
] table[
x = datetime,
y = knwn_mor_sim,
col sep=comma,
]{figdata/cloud_temporal.csv};
\addlegendentry{MoR}

\addplot[
dashed,
mycolor3,
line width=1.50,
] table[
x = datetime,
y = knwn_ros_sim,
col sep=comma,
]{figdata/cloud_temporal.csv};
\addlegendentry{RoS}

\addplot[
thick,
mycolor1,
line width=2,
] table[
x = datetime,
y = knwn_real_prop,
col sep=comma,
]{figdata/cloud_temporal.csv};
\addlegendentry{Full-crawl \cite{Balduf2023Cloud}}

% \addplot[
% thick,
% mycolor2,
% line width=1.50,
% ] table[
% x = datetime,
% y = knwn_ros_sim,
% col sep=comma,
% ]{figdata/cloud_temporal.csv};
% \addlegendentry{RoS}

% \addplot[
% thick,
% mycolor1,
% line width=1.50,
% ] table[
% x = datetime,
% y = knwn_real_prop,
% col sep=comma,
% ]{figdata/cloud_temporal.csv};
% \addlegendentry{\cite{Balduf2023Cloud}}

% \addlegendimage{area legend, fill=mycolor2!40}
%     \addlegendentry{Cloud}
% \addlegendimage{area legend, fill=mycolor4!40}
%     \addlegendentry{Non-cloud}

% \addplot[name path=lower2, fill=none, smooth, draw=none] table[x=datetime, y expr=0, col sep=comma]{figdata/cloud_temporal.csv};
% \addplot[name path=upper2, fill=none, smooth, draw=none] table[x=datetime, y=knwn_ros_sim, col sep=comma]{figdata/cloud_temporal.csv};
% \addplot [mycolor2!40] fill between [of=lower2 and upper2];

% \addplot[name path=lower2, fill=none, smooth, draw=none] table[x=datetime, y=knwn_ros_sim, col sep=comma]{figdata/cloud_temporal.csv};
% \addplot[name path=upper2, fill=none, smooth, draw=none] table[x=datetime, y expr=1, col sep=comma]{figdata/cloud_temporal.csv};
% \addplot [mycolor4!40] fill between [of=lower2 and upper2];

\end{axis}
\end{tikzpicture}
}
\end{minipage}
\end{tabular}
}
\vspace{-10pt}

%% file: figtex/sim_results.tex
\definecolor{mycolor1}{rgb}{0.00000,0.44700,0.74100}
\definecolor{mycolor2}{rgb}{0.85000,0.32500,0.09800}
\definecolor{mycolor3}{rgb}{0.92900,0.69400,0.12500}
\definecolor{mycolor4}{rgb}{0.46667,0.67451,0.18824}
\definecolor{mycolor5}{rgb}{0.49412,0.18431,0.30588}
\definecolor{mycolor6}{rgb}{0.30196,0.69020,0.93333}

\resizebox{\linewidth}{!}{
\begin{tabular}{c c c c c}
\hspace{-12pt}
\begin{minipage}{0.25\linewidth}
\resizebox{\linewidth}{!}{
\pgfplotstableread[col sep=comma]{figdata/country_stats_round_5553.csv}\datatable
\begin{tikzpicture}
\begin{axis}[
    ybar,
    bar width=2.30pt,
    scale = 2.50,
    width=\textwidth,
    height=\textwidth,
    xlabel={Country},
    ylabel={Attribute Rate},
    title = {(a) Country},
    xtick=data,
    xticklabels from table={\datatable}{country},
    x tick label style={
        rotate=45,
        anchor=east,
        font=\small
    },
    enlarge x limits=0.15,
    ymin=0,
]

\addplot+[
    fill=mycolor1!60,
    draw=mycolor1!80!black
    ] table[
    x expr=\coordindex,
    y=props_gt,
    col sep = comma
]{figdata/country_stats_round_5553.csv};

\addplot+[
    fill=mycolor2!60,
    draw=mycolor2!80!black
    ] table[
    x expr=\coordindex,
    y=mor_bckt,
    col sep = comma
]{figdata/country_stats_round_5553.csv};

\addplot+[
    fill=mycolor3!60,
    draw=mycolor3!80!black
    ] table[
    x expr=\coordindex,
    y=ros_bckt,
    col sep = comma
]{figdata/country_stats_round_5553.csv};
\legend{Full-crawl \cite{Balduf2023Cloud}, MoR (10 nodes), RoS (10 nodes)}
\end{axis}
\end{tikzpicture}
}
\end{minipage}
&
\hspace{-12pt}
\begin{minipage}{0.25\linewidth}
\resizebox{\linewidth}{!}{
\begin{tikzpicture}
\begin{axis}[
    ybar,
    bar width=2.30pt,
    scale = 2.50,
    width=\textwidth,
    height=\textwidth,
    xlabel={Cloud},
    ylabel={Attribute Rate},
    title = {(b) Cloud Provider},
    xtick=data,
    xticklabels={NC, HO, AA, CH, OV, CG, DO, VU},
    x tick label style={
        rotate=45,
        anchor=east,
        font=\small
    },
    enlarge x limits=0.15,
    ymin=0,
]

\addplot+[
    fill=mycolor1!60,
    draw=mycolor1!80!black
    ] table[
    x expr=\coordindex,
    y=props_gt,
    col sep = comma
]{figdata/cloud_stats_round_5569.csv};

\addplot+[
    fill=mycolor2!60,
    draw=mycolor2!80!black
    ] table[
    x expr=\coordindex,
    y=mor_bckt,
    col sep = comma
]{figdata/cloud_stats_round_5569.csv};

\addplot+[
    fill=mycolor3!60,
    draw=mycolor3!80!black
    ] table[
    x expr=\coordindex,
    y=ros_bckt,
    col sep = comma
]{figdata/cloud_stats_round_5569.csv};

\legend{Full-crawl \cite{Balduf2023Cloud}, MoR (10 nodes), RoS (10 nodes)}

\end{axis}
\end{tikzpicture}
}
\end{minipage}
&
\hspace{-12pt}
\begin{minipage}{0.25\linewidth}
\resizebox{\linewidth}{!}{
\begin{tikzpicture}
\begin{axis}[
    scale = 2.50,
    width=\textwidth,
    height=\textwidth,
    legend pos = south east,
    xlabel={Date},
    ylabel={Attribute Rate},
    title = {(c) US},
    date coordinates in=x,
    xticklabel={\day/\month},
    x tick label style={
        rotate=45,
        anchor=east,
        font=\small
    },
    enlarge x limits=0.025,
    ymin=0.22,
    ymax=0.40,
    scaled y ticks=base 10:2,
    ytick scale label code/.code={$\,\times 10^{-2}$},
]

\addplot[
smooth,
thick,
mycolor2,
line width=1.50,
] table[
x = datetime,
y = mor_bckt,
col sep=comma,
]{figdata/country_temporal_bckt2.csv};
\addlegendentry{MoR (10 nodes)}

\addplot[
smooth,
dashed,
mycolor3,
line width=1.50,
] table[
x = datetime,
y = ros_bckt,
col sep=comma,
]{figdata/country_temporal_bckt2.csv};
\addlegendentry{RoS (10 nodes)}

\addplot[
smooth,
thick,
mycolor1,
line width=2,
] table[
x = datetime,
y = real_prop,
col sep=comma,
]{figdata/country_temporal_bckt.csv};
\addlegendentry{Full-crawl \cite{Balduf2023Cloud}}

\end{axis}
\end{tikzpicture}
}
\end{minipage}
&
\hspace{-12pt}
\begin{minipage}{0.25\linewidth}
\resizebox{\linewidth}{!}{
\begin{tikzpicture}
\begin{axis}[
    scale = 2.50,
    width=\textwidth,
    height=\textwidth,
    legend pos = south east,
    xlabel={Date},
    ylabel={Attribute Rate},
    title = {(d) DE},
    date coordinates in=x,
    xticklabel={\day/\month},
    x tick label style={
        rotate=45,
        anchor=east,
        font=\small
    },
    enlarge x limits=0.025,
    ymin=0.06,
    ymax=0.15,
    scaled y ticks=base 10:2,
    ytick scale label code/.code={$\,\times 10^{-2}$},
]

\addplot[
smooth,
thick,
mycolor2,
line width=1.50,
] table[
x = datetime,
y = de_mor_bckt,
col sep=comma,
]{figdata/country_temporal_bckt2.csv};
\addlegendentry{MoR (10 nodes)}

\addplot[
smooth,
dashed,
mycolor3,
line width=1.50,
] table[
x = datetime,
y = de_ros_bckt,
col sep=comma,
]{figdata/country_temporal_bckt2.csv};
\addlegendentry{RoS (10 nodes)}

\addplot[
smooth,
thick,
mycolor1,
line width=2,
] table[
x = datetime,
y = de_real_prop,
col sep=comma,
]{figdata/country_temporal_bckt.csv};
\addlegendentry{Full-crawl \cite{Balduf2023Cloud}}

\end{axis}
\end{tikzpicture}
}
\end{minipage}
&
\hspace{-12pt}
\begin{minipage}{0.25\linewidth}
\resizebox{\linewidth}{!}{
\begin{tikzpicture}
\begin{axis}[
    scale = 2.50,
    width=\textwidth,
    height=\textwidth,
    legend pos = south east,
    xlabel={Date},
    ylabel={Attribute Rate},
    title = {(e) CA},
    date coordinates in=x,
    xticklabel={\day/\month},
    x tick label style={
        rotate=45,
        anchor=east,
        font=\small
    },
    enlarge x limits=0.025,
    ymin=0.00,
    ymax=0.06,
    scaled y ticks=base 10:2,
    ytick scale label code/.code={$\,\times 10^{-2}$},
]

\addplot[
smooth,
thick,
mycolor2,
line width=1.50,
] table[
x = datetime,
y = ca_mor_bckt,
col sep=comma,
]{figdata/country_temporal_bckt2.csv};
\addlegendentry{MoR (10 nodes)}

\addplot[
smooth,
dashed,
mycolor3,
line width=1.50,
] table[
x = datetime,
y = ca_ros_bckt,
col sep=comma,
]{figdata/country_temporal_bckt2.csv};
\addlegendentry{RoS (10 nodes)}

\addplot[
smooth,
thick,
mycolor1,
line width=2,
] table[
x = datetime,
y = ca_real_prop,
col sep=comma,
]{figdata/country_temporal_bckt.csv};
\addlegendentry{Full-crawl \cite{Balduf2023Cloud}}

\end{axis}
\end{tikzpicture}
}
\end{minipage}
\\
\hspace{-12pt}
\begin{minipage}{0.25\linewidth}
\resizebox{\linewidth}{!}{
\begin{tikzpicture}
\begin{axis}[
    scale = 2.50,
    width=\textwidth,
    height=\textwidth,
    legend pos = south east,
    xlabel={Date},
    ylabel={Attribute Rate},
    title = {(f) Hetzner Online},
    date coordinates in=x,
    xticklabel={\day/\month},
    x tick label style={
        rotate=45,
        anchor=east,
        font=\small
    },
    enlarge x limits=0.025,
    ymin=0.06,
    ymax=0.12,
    scaled y ticks=base 10:2,
    ytick scale label code/.code={$\,\times 10^{-2}$},
]

\addplot[
smooth,
thick,
mycolor2,
line width=1.50,
] table[
x = datetime,
y = mor_bckt,
col sep=comma,
]{figdata/cloud_temporal_bckt2.csv};
\addlegendentry{MoR (10 nodes)}

\addplot[
smooth,
dashed,
mycolor3,
line width=1.50,
] table[
x = datetime,
y = ros_bckt,
col sep=comma,
]{figdata/cloud_temporal_bckt2.csv};
\addlegendentry{RoS (10 nodes)}

\addplot[
smooth,
thick,
mycolor1,
line width=2,
] table[
x = datetime,
y = ho_real_prop,
col sep=comma,
]{figdata/cloud_temporal_bckt.csv};
\addlegendentry{Full-crawl \cite{Balduf2023Cloud}}

\end{axis}
\end{tikzpicture}
}
\end{minipage}
&
\hspace{-12pt}
\begin{minipage}{0.25\linewidth}
\resizebox{\linewidth}{!}{
\begin{tikzpicture}
\begin{axis}[
    scale = 2.50,
    width=\textwidth,
    height=\textwidth,
    legend pos = south east,
    xlabel={Date},
    ylabel={Attribute Rate},
    title = {(g) Amazon AWS},
    date coordinates in=x,
    xticklabel={\day/\month},
    x tick label style={
        rotate=45,
        anchor=east,
        font=\small
    },
    enlarge x limits=0.025,
    ymin=0.02,
    ymax=0.12,
    scaled y ticks=base 10:2,
    ytick scale label code/.code={$\,\times 10^{-2}$},
]

\addplot[
smooth,
thick,
mycolor2,
line width=1.50,
] table[
x = datetime,
y = aa_mor_bckt,
col sep=comma,
]{figdata/cloud_temporal_bckt2.csv};
\addlegendentry{MoR (10 nodes)}

\addplot[
smooth,
dashed,
mycolor3,
line width=1.50,
] table[
x = datetime,
y = aa_ros_bckt,
col sep=comma,
]{figdata/cloud_temporal_bckt2.csv};
\addlegendentry{RoS (10 nodes)}

\addplot[
smooth,
thick,
mycolor1,
line width=2,
] table[
x = datetime,
y = aa_real_prop,
col sep=comma,
]{figdata/cloud_temporal_bckt.csv};
\addlegendentry{Full-crawl \cite{Balduf2023Cloud}}

\end{axis}
\end{tikzpicture}
}
\end{minipage}
&
\hspace{-12pt}
\begin{minipage}{0.25\linewidth}
\resizebox{\linewidth}{!}{
\begin{tikzpicture}
\begin{axis}[
    scale = 2.50,
    width=\textwidth,
    height=\textwidth,
    legend pos = south east,
    xlabel={Date},
    ylabel={Attribute Rate},
    title = {(h) Digital Ocean},
    date coordinates in=x,
    xticklabel={\day/\month},
    x tick label style={
        rotate=45,
        anchor=east,
        font=\small
    },
    enlarge x limits=0.025,
    ymin=0.02,
    ymax=0.07,
    scaled y ticks=base 10:2,
    ytick scale label code/.code={$\,\times 10^{-2}$},
]

\addplot[
smooth,
thick,
mycolor2,
line width=1.50,
] table[
x = datetime,
y = do_mor_bckt,
col sep=comma,
]{figdata/cloud_temporal_bckt2.csv};
\addlegendentry{MoR (10 nodes)}

\addplot[
smooth,
dashed,
mycolor3,
line width=1.50,
] table[
x = datetime,
y = do_ros_bckt,
col sep=comma,
]{figdata/cloud_temporal_bckt2.csv};
\addlegendentry{RoS (10 nodes)}

\addplot[
smooth,
thick,
mycolor1,
line width=2,
] table[
x = datetime,
y = do_real_prop,
col sep=comma,
]{figdata/cloud_temporal_bckt.csv};
\addlegendentry{Full-crawl \cite{Balduf2023Cloud}}

\end{axis}
\end{tikzpicture}
}
\end{minipage}
&
\hspace{-12pt}
\begin{minipage}{0.25\linewidth}
\resizebox{\linewidth}{!}{
\begin{tikzpicture}
\begin{axis}[
    scale = 2.50,
    width=\textwidth,
    height=\textwidth,
    legend pos = south east,
    xlabel={Date},
    ylabel={Attribute Rate},
    title = {(i) Non-cloud},
    date coordinates in=x,
    xticklabel={\day/\month},
    x tick label style={
        rotate=45,
        anchor=east,
        font=\small
    },
    enlarge x limits=0.025,
    ymin=0.22,
    ymax=0.40,
    scaled y ticks=base 10:2,
    ytick scale label code/.code={$\,\times 10^{-2}$},
]

\addplot[
smooth,
thick,
mycolor2,
line width=1.50,
] table[
x = datetime,
y = un_mor_bckt,
col sep=comma,
]{figdata/cloud_temporal_bckt2.csv};
\addlegendentry{MoR (10 nodes)}

\addplot[
smooth,
dashed,
mycolor3,
line width=1.50,
] table[
x = datetime,
y = un_ros_bckt,
col sep=comma,
]{figdata/cloud_temporal_bckt2.csv};
\addlegendentry{RoS (10 nodes)}

\addplot[
smooth,
thick,
mycolor1,
line width=2,
] table[
x = datetime,
y = un_real_prop,
col sep=comma,
]{figdata/cloud_temporal_bckt.csv};
\addlegendentry{Full-crawl \cite{Balduf2023Cloud}}

% \addplot[
% smooth,
% thick,
% mycolor2,
% line width=1.50,
% ] table[
% x = datetime,
% y = kn_mor_bckt,
% col sep=comma,
% ]{figdata/cloud_temporal_bckt2.csv};
% \addlegendentry{MoR (10 nodes)}

% \addplot[
% smooth,
% thick,
% mycolor1,
% line width=1.50,
% ] table[
% x = datetime,
% y = kn_real_prop,
% col sep=comma,
% ]{figdata/cloud_temporal_bckt.csv};
% \addlegendentry{\cite{Balduf2023Cloud}}

% \addlegendimage{area legend, fill=mycolor2!40}
%     \addlegendentry{Cloud}
% \addlegendimage{area legend, fill=mycolor4!40}
%     \addlegendentry{Non-cloud}

% \addplot[name path=lower2, fill=none, smooth, draw=none] table[x=datetime, y expr=0, col sep=comma]{figdata/cloud_temporal_bckt2.csv};
% \addplot[name path=upper2, fill=none, smooth, draw=none] table[x=datetime, y=kn_mor_bckt, col sep=comma]{figdata/cloud_temporal_bckt2.csv};
% \addplot [mycolor2!40] fill between [of=lower2 and upper2];

% \addplot[name path=lower2, fill=none, smooth, draw=none] table[x=datetime, y=kn_mor_bckt, col sep=comma]{figdata/cloud_temporal_bckt2.csv};
% \addplot[name path=upper2, fill=none, smooth, draw=none] table[x=datetime, y expr=1, col sep=comma]{figdata/cloud_temporal_bckt2.csv};
% \addplot [mycolor4!40] fill between [of=lower2 and upper2];

\end{axis}
\end{tikzpicture}
}
\end{minipage}
&
\hspace{-12pt}
\begin{minipage}{0.25\linewidth}
\resizebox{\linewidth}{!}{
\begin{tikzpicture}
\begin{axis}[
    scale = 2.50,
    width=\textwidth,
    height=\textwidth,
    legend pos = south east,
    xlabel={Date},
    ylabel={Attribute Rate},
    title = {(j) Cloud},
    date coordinates in=x,
    xticklabel={\day/\month},
    x tick label style={
        rotate=45,
        anchor=east,
        font=\small
    },
    enlarge x limits=0.025,
    ymin=0.56,
    ymax=0.74,
    scaled y ticks=base 10:2,
    ytick scale label code/.code={$\,\times 10^{-2}$},
]

\addplot[
smooth,
thick,
mycolor2,
line width=1.50,
] table[
x = datetime,
y = kn_mor_bckt,
col sep=comma,
]{figdata/cloud_temporal_bckt2.csv};
\addlegendentry{MoR (10 nodes)}

\addplot[
smooth,
dashed,
mycolor3,
line width=1.50,
] table[
x = datetime,
y = kn_ros_bckt,
col sep=comma,
]{figdata/cloud_temporal_bckt2.csv};
\addlegendentry{RoS (10 nodes)}

\addplot[
smooth,
thick,
mycolor1,
line width=2,
] table[
x = datetime,
y = kn_real_prop,
col sep=comma,
]{figdata/cloud_temporal_bckt.csv};
\addlegendentry{Full-crawl \cite{Balduf2023Cloud}}

% \addplot[
% smooth,
% thick,
% mycolor2,
% line width=1.50,
% ] table[
% x = datetime,
% y = kn_ros_bckt,
% col sep=comma,
% ]{figdata/cloud_temporal_bckt2.csv};
% \addlegendentry{RoS (10 nodes)}

% \addplot[
% smooth,
% thick,
% mycolor1,
% line width=1.50,
% ] table[
% x = datetime,
% y = kn_real_prop,
% col sep=comma,
% ]{figdata/cloud_temporal_bckt.csv};
% \addlegendentry{\cite{Balduf2023Cloud}}

% \addlegendimage{area legend, fill=mycolor2!40}
%     \addlegendentry{Cloud}
% \addlegendimage{area legend, fill=mycolor4!40}
%     \addlegendentry{Non-cloud}

% \addplot[name path=lower2, fill=none, smooth, draw=none] table[x=datetime, y expr=0, col sep=comma]{figdata/cloud_temporal_bckt2.csv};
% \addplot[name path=upper2, fill=none, smooth, draw=none] table[x=datetime, y=kn_mor_bckt, col sep=comma]{figdata/cloud_temporal_bckt2.csv};
% \addplot [mycolor2!40] fill between [of=lower2 and upper2];

% \addplot[name path=lower2, fill=none, smooth, draw=none] table[x=datetime, y=kn_mor_bckt, col sep=comma]{figdata/cloud_temporal_bckt2.csv};
% \addplot[name path=upper2, fill=none, smooth, draw=none] table[x=datetime, y expr=1, col sep=comma]{figdata/cloud_temporal_bckt2.csv};
% \addplot [mycolor4!40] fill between [of=lower2 and upper2];

\end{axis}
\end{tikzpicture}
}
\end{minipage}
\end{tabular}
}
\vspace{-10pt}

%% file: figtex/nonat_results.tex
\definecolor{mycolor1}{rgb}{0.00000,0.44700,0.74100}
\definecolor{mycolor2}{rgb}{0.85000,0.32500,0.09800}
\definecolor{mycolor3}{rgb}{0.92900,0.69400,0.12500}
\definecolor{mycolor4}{rgb}{0.46667,0.67451,0.18824}
\definecolor{mycolor5}{rgb}{0.49412,0.18431,0.30588}
\definecolor{mycolor6}{rgb}{0.30196,0.69020,0.93333}

\resizebox{.9\linewidth}{!}{
\begin{tabular}{c c c c}
\hspace{-12pt}
\begin{minipage}{0.25\linewidth}
\resizebox{\linewidth}{!}{
\pgfplotstableread[col sep=comma]{figdata/country_stats_round_5977_NONAT.csv}\datatable
\begin{tikzpicture}
\begin{axis}[
    ybar,
    bar width=2.30pt,
    scale = 2.50,
    width=\textwidth,
    height=\textwidth,
    xlabel={Country},
    ylabel={Attribute Rate},
    title = {(a) Country},
    xtick=data,
    xticklabels from table={\datatable}{country},
    x tick label style={
        rotate=45,
        anchor=east,
        font=\small
    },
    enlarge x limits=0.15,
    ymin=0,
]

\addplot+[
    fill=mycolor1!60,
    draw=mycolor1!80!black
    ] table[
    x expr=\coordindex,
    y=props_gt,
    col sep = comma
]{figdata/country_stats_round_5977_NONAT.csv};

\addplot+[
    fill=mycolor2!60,
    draw=mycolor2!80!black
    ] table[
    x expr=\coordindex,
    y=mor_bckt,
    col sep = comma
]{figdata/country_stats_round_5977_NONAT.csv};

\addplot+[
    fill=mycolor3!60,
    draw=mycolor3!80!black
    ] table[
    x expr=\coordindex,
    y=ros_bckt,
    col sep = comma
]{figdata/country_stats_round_5977_NONAT.csv};
\legend{Full-crawl \cite{Balduf2023Cloud}, MoR (5 NATed), RoS (5 NATed)}
\end{axis}
\end{tikzpicture}
}
\end{minipage}
&
\hspace{-12pt}
\begin{minipage}{0.25\linewidth}
\resizebox{\linewidth}{!}{
\begin{tikzpicture}
\begin{axis}[
    ybar,
    bar width=2.30pt,
    scale = 2.50,
    width=\textwidth,
    height=\textwidth,
    xlabel={Cloud},
    ylabel={Attribute Rate},
    title = {(b) Cloud Provider},
    xtick=data,
    xticklabels={UN, HO, AA, CH, OV, CG, DO, VU},
    x tick label style={
        rotate=45,
        anchor=east,
        font=\small
    },
    enlarge x limits=0.15,
    ymin=0,
]

\addplot+[
    fill=mycolor1!60,
    draw=mycolor1!80!black
    ] table[
    x expr=\coordindex,
    y=props_gt,
    col sep = comma
]{figdata/cloud_stats_round_5929_NONAT.csv};

\addplot+[
    fill=mycolor2!60,
    draw=mycolor2!80!black
    ] table[
    x expr=\coordindex,
    y=mor_bckt,
    col sep = comma
]{figdata/cloud_stats_round_5929_NONAT.csv};

\addplot+[
    fill=mycolor3!60,
    draw=mycolor3!80!black
    ] table[
    x expr=\coordindex,
    y=ros_bckt,
    col sep = comma
]{figdata/cloud_stats_round_5929_NONAT.csv};

\legend{Full-crawl \cite{Balduf2023Cloud}, MoR (5 NATed), RoS (5 NATed)}

\end{axis}
\end{tikzpicture}
}
\end{minipage}
&
\hspace{-12pt}
\begin{minipage}{0.25\linewidth}
\resizebox{\linewidth}{!}{
\begin{tikzpicture}
\begin{axis}[
    scale = 2.50,
    width=\textwidth,
    height=\textwidth,
    legend pos = south east,
    xlabel={Date},
    ylabel={Attribute Rate},
    title = {(c) US},
    date coordinates in=x,
    xticklabel={\day/\month},
    x tick label style={
        rotate=45,
        anchor=east,
        font=\small
    },
    enlarge x limits=0.025,
    ymin=0.25,
    ymax=0.40,
    scaled y ticks=base 10:2,
    ytick scale label code/.code={$\,\times 10^{-2}$},
]

\addplot[
smooth,
thick,
mycolor2,
line width=1.50,
] table[
x = datetime,
y = mor_bckt,
col sep=comma,
]{figdata/country_temporal_bckt2_NONAT.csv};
\addlegendentry{MoR (5 NATed)}

\addplot[
smooth,
dashed,
mycolor3,
line width=1.50,
] table[
x = datetime,
y = ros_bckt,
col sep=comma,
]{figdata/country_temporal_bckt2_NONAT.csv};
\addlegendentry{RoS (5 NATed)}

\addplot[
smooth,
thick,
mycolor1,
line width=2,
] table[
x = datetime,
y = real_prop,
col sep=comma,
]{figdata/country_temporal_bckt.csv};
\addlegendentry{Full-crawl \cite{Balduf2023Cloud}}

\end{axis}
\end{tikzpicture}
}
\end{minipage}
&
\hspace{-12pt}
\begin{minipage}{0.25\linewidth}
\resizebox{\linewidth}{!}{
\begin{tikzpicture}
\begin{axis}[
    scale = 2.50,
    width=\textwidth,
    height=\textwidth,
    legend pos = south east,
    xlabel={Date},
    ylabel={Attribute Rate},
    title = {(d) DE},
    date coordinates in=x,
    xticklabel={\day/\month},
    x tick label style={
        rotate=45,
        anchor=east,
        font=\small
    },
    enlarge x limits=0.025,
    ymin=0.06,
    ymax=0.15,
    scaled y ticks=base 10:2,
    ytick scale label code/.code={$\,\times 10^{-2}$},
]

\addplot[
smooth,
thick,
mycolor2,
line width=1.50,
] table[
x = datetime,
y = de_mor_bckt,
col sep=comma,
]{figdata/country_temporal_bckt2_NONAT.csv};
\addlegendentry{MoR (5 NATed)}

\addplot[
smooth,
dashed,
mycolor3,
line width=1.50,
] table[
x = datetime,
y = de_ros_bckt,
col sep=comma,
]{figdata/country_temporal_bckt2_NONAT.csv};
\addlegendentry{RoS (5 NATed)}

\addplot[
smooth,
thick,
mycolor1,
line width=2,
] table[
x = datetime,
y = de_real_prop,
col sep=comma,
]{figdata/country_temporal_bckt.csv};
\addlegendentry{Full-crawl \cite{Balduf2023Cloud}}

\end{axis}
\end{tikzpicture}
}
\end{minipage}
\\
\hspace{-12pt}
\begin{minipage}{0.25\linewidth}
\resizebox{\linewidth}{!}{
\begin{tikzpicture}
\begin{axis}[
    scale = 2.50,
    width=\textwidth,
    height=\textwidth,
    legend pos = south east,
    xlabel={Date},
    ylabel={Attribute Rate},
    title = {(e) CA},
    date coordinates in=x,
    xticklabel={\day/\month},
    x tick label style={
        rotate=45,
        anchor=east,
        font=\small
    },
    enlarge x limits=0.025,
    ymin=0.00,
    ymax=0.05,
    scaled y ticks=base 10:2,
    ytick scale label code/.code={$\,\times 10^{-2}$},
]

\addplot[
smooth,
thick,
mycolor2,
line width=1.50,
] table[
x = datetime,
y = ca_mor_bckt,
col sep=comma,
]{figdata/country_temporal_bckt2_NONAT.csv};
\addlegendentry{MoR (5 NATed)}

\addplot[
smooth,
dashed,
mycolor3,
line width=1.50,
] table[
x = datetime,
y = ca_ros_bckt,
col sep=comma,
]{figdata/country_temporal_bckt2_NONAT.csv};
\addlegendentry{RoS (5 NATed)}

\addplot[
smooth,
thick,
mycolor1,
line width=2,
] table[
x = datetime,
y = ca_real_prop,
col sep=comma,
]{figdata/country_temporal_bckt.csv};
\addlegendentry{Full-crawl \cite{Balduf2023Cloud}}

\end{axis}
\end{tikzpicture}
}
\end{minipage}
&
\hspace{-12pt}
\begin{minipage}{0.25\linewidth}
\resizebox{\linewidth}{!}{
\begin{tikzpicture}
\begin{axis}[
    scale = 2.50,
    width=\textwidth,
    height=\textwidth,
    legend pos = south east,
    xlabel={Date},
    ylabel={Attribute Rate},
    title = {(f) Hetzner Online},
    date coordinates in=x,
    xticklabel={\day/\month},
    x tick label style={
        rotate=45,
        anchor=east,
        font=\small
    },
    enlarge x limits=0.025,
    ymin=0.04,
    ymax=0.14,
    scaled y ticks=base 10:2,
    ytick scale label code/.code={$\,\times 10^{-2}$},
]

\addplot[
smooth,
thick,
mycolor2,
line width=1.50,
] table[
x = datetime,
y = mor_bckt,
col sep=comma,
]{figdata/cloud_temporal_bckt2_NONAT.csv};
\addlegendentry{MoR (5 NATed)}

\addplot[
smooth,
dashed,
mycolor3,
line width=1.50,
] table[
x = datetime,
y = ros_bckt,
col sep=comma,
]{figdata/cloud_temporal_bckt2_NONAT.csv};
\addlegendentry{RoS (5 NATed)}

\addplot[
smooth,
thick,
mycolor1,
line width=2,
] table[
x = datetime,
y = ho_real_prop,
col sep=comma,
]{figdata/cloud_temporal_bckt.csv};
\addlegendentry{Full-crawl \cite{Balduf2023Cloud}}

\end{axis}
\end{tikzpicture}
}
\end{minipage}
&
\hspace{-12pt}
\begin{minipage}{0.25\linewidth}
\resizebox{\linewidth}{!}{
\begin{tikzpicture}
\begin{axis}[
    scale = 2.50,
    width=\textwidth,
    height=\textwidth,
    legend pos = south east,
    xlabel={Date},
    ylabel={Attribute Rate},
    title = {(g) Amazon AWS},
    date coordinates in=x,
    xticklabel={\day/\month},
    x tick label style={
        rotate=45,
        anchor=east,
        font=\small
    },
    enlarge x limits=0.025,
    ymin=0.02,
    ymax=0.15,
    scaled y ticks=base 10:2,
    ytick scale label code/.code={$\,\times 10^{-2}$},
]

\addplot[
smooth,
thick,
mycolor2,
line width=1.50,
] table[
x = datetime,
y = aa_mor_bckt,
col sep=comma,
]{figdata/cloud_temporal_bckt2_NONAT.csv};
\addlegendentry{MoR (5 NATed)}

\addplot[
smooth,
dashed,
mycolor3,
line width=1.50,
] table[
x = datetime,
y = aa_ros_bckt,
col sep=comma,
]{figdata/cloud_temporal_bckt2_NONAT.csv};
\addlegendentry{RoS (5 NATed)}

\addplot[
smooth,
thick,
mycolor1,
line width=2,
] table[
x = datetime,
y = aa_real_prop,
col sep=comma,
]{figdata/cloud_temporal_bckt.csv};
\addlegendentry{Full-crawl \cite{Balduf2023Cloud}}

\end{axis}
\end{tikzpicture}
}
\end{minipage}
&
\hspace{-12pt}
\begin{minipage}{0.25\linewidth}
\resizebox{\linewidth}{!}{
\begin{tikzpicture}
\begin{axis}[
    scale = 2.50,
    width=\textwidth,
    height=\textwidth,
    legend pos = south east,
    xlabel={Date},
    ylabel={Attribute Rate},
    title = {(h) Digital Ocean},
    date coordinates in=x,
    xticklabel={\day/\month},
    x tick label style={
        rotate=45,
        anchor=east,
        font=\small
    },
    enlarge x limits=0.025,
    ymin=0.00,
    ymax=0.08,
    scaled y ticks=base 10:2,
    ytick scale label code/.code={$\,\times 10^{-2}$},
]

\addplot[
smooth,
thick,
mycolor2,
line width=1.50,
] table[
x = datetime,
y = do_mor_bckt,
col sep=comma,
]{figdata/cloud_temporal_bckt2_NONAT.csv};
\addlegendentry{MoR (5 NATed)}

\addplot[
smooth,
dashed,
mycolor3,
line width=1.50,
] table[
x = datetime,
y = do_ros_bckt,
col sep=comma,
]{figdata/cloud_temporal_bckt2_NONAT.csv};
\addlegendentry{RoS (5 NATed)}

\addplot[
smooth,
thick,
mycolor1,
line width=2,
] table[
x = datetime,
y = do_real_prop,
col sep=comma,
]{figdata/cloud_temporal_bckt.csv};
\addlegendentry{Full-crawl \cite{Balduf2023Cloud}}

\end{axis}
\end{tikzpicture}
}
\end{minipage}
\end{tabular}
}
\vspace{-10pt}

%% file: figtex/nat_results.tex
\definecolor{mycolor1}{rgb}{0.00000,0.44700,0.74100}
\definecolor{mycolor2}{rgb}{0.85000,0.32500,0.09800}
\definecolor{mycolor3}{rgb}{0.92900,0.69400,0.12500}
\definecolor{mycolor4}{rgb}{0.46667,0.67451,0.18824}
\definecolor{mycolor5}{rgb}{0.49412,0.18431,0.30588}
\definecolor{mycolor6}{rgb}{0.30196,0.69020,0.93333}

\resizebox{.9\linewidth}{!}{
\begin{tabular}{c c c c}
\hspace{-12pt}
\begin{minipage}{0.25\linewidth}
\resizebox{\linewidth}{!}{
\pgfplotstableread[col sep=comma]{figdata/country_stats_round_5977_NAT.csv}\datatable
\begin{tikzpicture}
\begin{axis}[
    ybar,
    bar width=2.30pt,
    scale = 2.50,
    width=\textwidth,
    height=\textwidth,
    xlabel={Country},
    ylabel={Attribute Rate},
    title = {(a) Country},
    xtick=data,
    xticklabels from table={\datatable}{country},
    x tick label style={
        rotate=45,
        anchor=east,
        font=\small
    },
    enlarge x limits=0.15,
    ymin=0,
]

\addplot+[
    fill=mycolor1!60,
    draw=mycolor1!80!black
    ] table[
    x expr=\coordindex,
    y=props_gt,
    col sep = comma
]{figdata/country_stats_round_5977_NAT.csv};

\addplot+[
    fill=mycolor2!60,
    draw=mycolor2!80!black
    ] table[
    x expr=\coordindex,
    y=mor_bckt,
    col sep = comma
]{figdata/country_stats_round_5977_NAT.csv};

\addplot+[
    fill=mycolor3!60,
    draw=mycolor3!80!black
    ] table[
    x expr=\coordindex,
    y=ros_bckt,
    col sep = comma
]{figdata/country_stats_round_5977_NAT.csv};
\legend{Full-crawl \cite{Balduf2023Cloud}, MoR (5 NoNATed), RoS (5 NoNATed)}
\end{axis}
\end{tikzpicture}
}
\end{minipage}
&
\hspace{-12pt}
\begin{minipage}{0.25\linewidth}
\resizebox{\linewidth}{!}{
\begin{tikzpicture}
\begin{axis}[
    ybar,
    bar width=2.30pt,
    scale = 2.50,
    width=\textwidth,
    height=\textwidth,
    xlabel={Cloud},
    ylabel={Attribute Rate},
    title = {(b) Cloud Provider},
    xtick=data,
    xticklabels={UN, HO, AA, CH, OV, CG, DO, VU},
    x tick label style={
        rotate=45,
        anchor=east,
        font=\small
    },
    enlarge x limits=0.15,
    ymin=0,
]

\addplot+[
    fill=mycolor1!60,
    draw=mycolor1!80!black
    ] table[
    x expr=\coordindex,
    y=props_gt,
    col sep = comma
]{figdata/cloud_stats_round_5929_NAT.csv};

\addplot+[
    fill=mycolor2!60,
    draw=mycolor2!80!black
    ] table[
    x expr=\coordindex,
    y=mor_bckt,
    col sep = comma
]{figdata/cloud_stats_round_5929_NAT.csv};

\addplot+[
    fill=mycolor3!60,
    draw=mycolor3!80!black
    ] table[
    x expr=\coordindex,
    y=ros_bckt,
    col sep = comma
]{figdata/cloud_stats_round_5929_NAT.csv};

\legend{Full-crawl \cite{Balduf2023Cloud}, MoR (5 NoNATed), RoS (5 NoNATed)}

\end{axis}
\end{tikzpicture}
}
\end{minipage}
&
\hspace{-12pt}
\begin{minipage}{0.25\linewidth}
\resizebox{\linewidth}{!}{
\begin{tikzpicture}
\begin{axis}[
    scale = 2.50,
    width=\textwidth,
    height=\textwidth,
    legend pos = south east,
    xlabel={Date},
    ylabel={Attribute Rate},
    title = {(c) US},
    date coordinates in=x,
    xticklabel={\day/\month},
    x tick label style={
        rotate=45,
        anchor=east,
        font=\small
    },
    enlarge x limits=0.025,
    ymin=0.22,
    ymax=0.40,
    scaled y ticks=base 10:2,
    ytick scale label code/.code={$\,\times 10^{-2}$},
]

\addplot[
smooth,
thick,
mycolor2,
line width=1.50,
] table[
x = datetime,
y = mor_bckt,
col sep=comma,
]{figdata/country_temporal_bckt2_NAT.csv};
\addlegendentry{MoR (5 NoNATed)}

\addplot[
smooth,
dashed,
mycolor3,
line width=1.50,
] table[
x = datetime,
y = ros_bckt,
col sep=comma,
]{figdata/country_temporal_bckt2_NAT.csv};
\addlegendentry{RoS (5 NoNATed)}

\addplot[
smooth,
thick,
mycolor1,
line width=2,
] table[
x = datetime,
y = real_prop,
col sep=comma,
]{figdata/country_temporal_bckt.csv};
\addlegendentry{Full-crawl \cite{Balduf2023Cloud}}

\end{axis}
\end{tikzpicture}
}
\end{minipage}
&
\hspace{-12pt}
\begin{minipage}{0.25\linewidth}
\resizebox{\linewidth}{!}{
\begin{tikzpicture}
\begin{axis}[
    scale = 2.50,
    width=\textwidth,
    height=\textwidth,
    legend pos = south east,
    xlabel={Date},
    ylabel={Attribute Rate},
    title = {(d) DE},
    date coordinates in=x,
    xticklabel={\day/\month},
    x tick label style={
        rotate=45,
        anchor=east,
        font=\small
    },
    enlarge x limits=0.025,
    ymin=0.06,
    ymax=0.16,
    scaled y ticks=base 10:2,
    ytick scale label code/.code={$\,\times 10^{-2}$},
]

\addplot[
smooth,
thick,
mycolor2,
line width=1.50,
] table[
x = datetime,
y = de_mor_bckt,
col sep=comma,
]{figdata/country_temporal_bckt2_NAT.csv};
\addlegendentry{MoR (5 NoNATed)}

\addplot[
smooth,
dashed,
mycolor3,
line width=1.50,
] table[
x = datetime,
y = de_ros_bckt,
col sep=comma,
]{figdata/country_temporal_bckt2_NAT.csv};
\addlegendentry{RoS (5 NoNATed)}

\addplot[
smooth,
thick,
mycolor1,
line width=2,
] table[
x = datetime,
y = de_real_prop,
col sep=comma,
]{figdata/country_temporal_bckt.csv};
\addlegendentry{Full-crawl \cite{Balduf2023Cloud}}

\end{axis}
\end{tikzpicture}
}
\end{minipage}
\\
\hspace{-12pt}
\begin{minipage}{0.25\linewidth}
\resizebox{\linewidth}{!}{
\begin{tikzpicture}
\begin{axis}[
    scale = 2.50,
    width=\textwidth,
    height=\textwidth,
    legend pos = south east,
    xlabel={Date},
    ylabel={Attribute Rate},
    title = {(e) CA},
    date coordinates in=x,
    xticklabel={\day/\month},
    x tick label style={
        rotate=45,
        anchor=east,
        font=\small
    },
    enlarge x limits=0.025,
    ymin=0.00,
    ymax=0.08,
    scaled y ticks=base 10:2,
    ytick scale label code/.code={$\,\times 10^{-2}$},
]

\addplot[
smooth,
thick,
mycolor2,
line width=1.50,
] table[
x = datetime,
y = ca_mor_bckt,
col sep=comma,
]{figdata/country_temporal_bckt2_NAT.csv};
\addlegendentry{MoR (5 NoNATed)}

\addplot[
smooth,
dashed,
mycolor3,
line width=1.50,
] table[
x = datetime,
y = ca_ros_bckt,
col sep=comma,
]{figdata/country_temporal_bckt2_NAT.csv};
\addlegendentry{RoS (5 NoNATed)}

\addplot[
smooth,
thick,
mycolor1,
line width=2,
] table[
x = datetime,
y = ca_real_prop,
col sep=comma,
]{figdata/country_temporal_bckt.csv};
\addlegendentry{Full-crawl \cite{Balduf2023Cloud}}

\end{axis}
\end{tikzpicture}
}
\end{minipage}
&
\hspace{-12pt}
\begin{minipage}{0.25\linewidth}
\resizebox{\linewidth}{!}{
\begin{tikzpicture}
\begin{axis}[
    scale = 2.50,
    width=\textwidth,
    height=\textwidth,
    legend pos = south east,
    xlabel={Date},
    ylabel={Attribute Rate},
    title = {(f) Hetzner Online},
    date coordinates in=x,
    xticklabel={\day/\month},
    x tick label style={
        rotate=45,
        anchor=east,
        font=\small
    },
    enlarge x limits=0.025,
    ymin=0.04,
    ymax=0.14,
    scaled y ticks=base 10:2,
    ytick scale label code/.code={$\,\times 10^{-2}$},
]

\addplot[
smooth,
thick,
mycolor2,
line width=1.50,
] table[
x = datetime,
y = mor_bckt,
col sep=comma,
]{figdata/cloud_temporal_bckt2_NAT.csv};
\addlegendentry{MoR (5 NoNATed)}

\addplot[
smooth,
dashed,
mycolor3,
line width=1.50,
] table[
x = datetime,
y = ros_bckt,
col sep=comma,
]{figdata/cloud_temporal_bckt2_NAT.csv};
\addlegendentry{RoS (5 NoNATed)}

\addplot[
smooth,
thick,
mycolor1,
line width=2,
] table[
x = datetime,
y = ho_real_prop,
col sep=comma,
]{figdata/cloud_temporal_bckt.csv};
\addlegendentry{Full-crawl \cite{Balduf2023Cloud}}

\end{axis}
\end{tikzpicture}
}
\end{minipage}
&
\hspace{-12pt}
\begin{minipage}{0.25\linewidth}
\resizebox{\linewidth}{!}{
\begin{tikzpicture}
\begin{axis}[
    scale = 2.50,
    width=\textwidth,
    height=\textwidth,
    legend pos = south east,
    xlabel={Date},
    ylabel={Attribute Rate},
    title = {(g) Amazon AWS},
    date coordinates in=x,
    xticklabel={\day/\month},
    x tick label style={
        rotate=45,
        anchor=east,
        font=\small
    },
    enlarge x limits=0.025,
    ymin=0.02,
    ymax=0.12,
    scaled y ticks=base 10:2,
    ytick scale label code/.code={$\,\times 10^{-2}$},
]

\addplot[
smooth,
thick,
mycolor2,
line width=1.50,
] table[
x = datetime,
y = aa_mor_bckt,
col sep=comma,
]{figdata/cloud_temporal_bckt2_NAT.csv};
\addlegendentry{MoR (5 NoNATed)}

\addplot[
smooth,
dashed,
mycolor3,
line width=1.50,
] table[
x = datetime,
y = aa_ros_bckt,
col sep=comma,
]{figdata/cloud_temporal_bckt2_NAT.csv};
\addlegendentry{RoS (5 NoNATed)}

\addplot[
smooth,
thick,
mycolor1,
line width=2,
] table[
x = datetime,
y = aa_real_prop,
col sep=comma,
]{figdata/cloud_temporal_bckt.csv};
\addlegendentry{Full-crawl \cite{Balduf2023Cloud}}

\end{axis}
\end{tikzpicture}
}
\end{minipage}
&
\hspace{-12pt}
\begin{minipage}{0.25\linewidth}
\resizebox{\linewidth}{!}{
\begin{tikzpicture}
\begin{axis}[
    scale = 2.50,
    width=\textwidth,
    height=\textwidth,
    legend pos = south east,
    xlabel={Date},
    ylabel={Attribute Rate},
    title = {(h) Digital Ocean},
    date coordinates in=x,
    xticklabel={\day/\month},
    x tick label style={
        rotate=45,
        anchor=east,
        font=\small
    },
    enlarge x limits=0.025,
    ymin=0.00,
    ymax=0.08,
    scaled y ticks=base 10:2,
    ytick scale label code/.code={$\,\times 10^{-2}$},
]

\addplot[
smooth,
thick,
mycolor2,
line width=1.50,
] table[
x = datetime,
y = do_mor_bckt,
col sep=comma,
]{figdata/cloud_temporal_bckt2_NAT.csv};
\addlegendentry{MoR (5 NoNATed)}

\addplot[
smooth,
dashed,
mycolor3,
line width=1.50,
] table[
x = datetime,
y = do_ros_bckt,
col sep=comma,
]{figdata/cloud_temporal_bckt2_NAT.csv};
\addlegendentry{RoS (5 NoNATed)}

\addplot[
smooth,
thick,
mycolor1,
line width=2,
] table[
x = datetime,
y = do_real_prop,
col sep=comma,
]{figdata/cloud_temporal_bckt.csv};
\addlegendentry{Full-crawl \cite{Balduf2023Cloud}}

\end{axis}
\end{tikzpicture}
}
\end{minipage}
\end{tabular}
}
\vspace{-10pt}